\documentclass[apjl,twocolumn]{emulateapj_mod}

\newcommand{\simlt}{\lower.5ex\hbox{$\; \buildrel < \over \sim \;$}}
\newcommand{\simgt}{\lower.5ex\hbox{$\; \buildrel > \over \sim \;$}}

\newcommand{\be}{\begin{equation}}
\newcommand{\ba}{\begin{eqnarray}}
\newcommand{\ee}{\end{equation}}
\newcommand{\ea}{\end{eqnarray}}

\slugcomment{Accepted by ApJ}

\shorttitle{Early-type galaxies in the UDF/GRAPES survey}
\shortauthors{Pasquali et al.}
\begin{document}
\title{The Structure and Star Formation History of Early-Type Galaxies
	in the UDF/GRAPES Survey}
\author{A. Pasquali}
\affil{Institut f\"ur Astronomie, HPF, ETH, 8093 Z\"urich, Switzerland}
\email{pasquali@phys.ethz.ch}

\author{I. Ferreras}
\affil{Department of Physics and Astronomy, University College London,
  Gower St. London WC1E 6BT, England}
\email{ferreras@star.ucl.ac.uk}

\author{N. Panagia}
\affil{STScI, 3700 San Martin Drive, Baltimore, MD 21218, USA}
\email{panagia@stsci.edu}

\author{E. Daddi}
\affil{Spitzer Fellow, NOAO, P.O. Box 26732, Tucson, AZ 85726, USA}
\email{edaddi@noao.edu}

\author{S. Malhotra, J.E. Rhoads, N. Pirzkal} 
\affil{STScI, 3700 San Martin Drive, Baltimore, MD 21218, USA}
\email{san@stsci.edu, rhoads@stsci.edu, npirzkal@stsci.edu}

\author{R.A. Windhorst}
\affil{Department of Physics \& Astronomy, Arizona State University, P.O. Box
871504, Tempe, AZ 85287-1504, USA}
\email{Rogier.Windhorst@asu.edu}

\author{A.M. Koekemoer, L. Moustakas, C. Xu}
\affil{STScI, 3700 San Martin Drive, Baltimore, MD 21218, USA}
\email{koekemoe@stsci.edu, leonidas@stsci.edu, chunxu@stsci.edu}

\author{C. Gronwall}
\affil{Department of Astronomy, Pennsylvania State University, 525 Davey
Laboratory, University Park, PA 16802}
\email{caryl@astro.psu.edu}

\begin{abstract}
We present a two-pronged approach to the formation of early-type galaxies,
using a sample of 18 galaxies at $0.5\simlt z\simlt 1$ from the HST/ACS  
Ultra Deep Field and GRAPES surveys:
1) We combine slitless low resolution spectroscopy from the GRAPES dataset with 
simple models of galaxy formation to explore their star formation histories. 
2) We also perform an analysis of their surface brightness distribution 
with the unprecedented details provided by the ACS superb
angular resolution and photometric depth. Our spectroscopic analysis
reveals that their stellar populations are rather homogeneous in age
and metallicity and formed at redshifts $z_F\sim 2-5$. Evolving them 
passively, they become practically indistinguishable from ellipticals
at $z =$ 0. Also, their isophotal shapes appear very similar to those
observed for nearby ellipticals, in that the percentages of disky and boxy
galaxies at $z \sim$ 1 are close to the values measured at $z =$ 0. Moreover, 
we find that the isophotal structure of $z\sim$ 1  early-type galaxies
obeys the correlations already observed among nearby ellipticals, i.e.
disky ellipticals have generally higher characteristic ellipticities, and
boxy ellipticals have larger half-light radii and are brighter in the 
rest-frame B band. In this respect then, no significant structural differences
are seen for ellipticals between $z =$ 0 and 1. Exception can be possibly made for the 
{\it a$_3$/a} parameter, which is larger at $z \sim$ 1 than usually measured
at $z =$ 0. The {\it a$_3$/a} parameter measures the deviations from a pure
elliptical isophote which are not symmetric with respect to the galaxy center,
as in the case of dust features and most notably of clumps. Blue clumps have been detected 
in nearly 50$\%$ of the $z\sim$ 1  early-type galaxies; their photometry is
suggestive of young star clusters or dwarf irregulars if they are assumed to be
at the same redshift as their host galaxies. We speculate that these clumps may 
represent recent accretion episodes, and that they could be a way to produce blue
cores if their dynamical time is such for them to rapidly sink to the galaxy center.
%Based on similarities in stellar mass and metallicity, we also speculate that
%the $z\sim$ 1  early-type galaxies may be the {\it observable} descendants of
%Lyman-break galaxies at $z \sim$ 3.

\end{abstract}


%%%%%%%%%%%%%%%%%%%%%%%%%%%%%%%%%%%%%%%%%%%%%%%%
\section{Introduction}
Early-type galaxies form a remarkably homogeneous class of objects
with a tight color-magnitude relation and a well
defined Fundamental Plane. The small scatter in their colors (Bower,
Lucey \& Ellis 1992, Stanford, Eisenhardt \& Dickinson 1998,) and
mass-to-light (M/L) ratios (cf. Kelson et al. 2000) suggest that early-type 
galaxies formed the bulk of their stars at high redshift (i.e. $z_F >$ 2 - 3,
van Dokkum et al. 1998, Thomas et al. 2005). Since then,
early-type galaxies continued to grow in mass at different paces: the more 
massive objects ($>$ 10$^{11}$ M$_{\odot}$) appear to have increased their 
mass by less than 1$\%$ since $z \simeq$ 1, while the less massive
grew by 20 - 40$\%$ (cf. Chen et al. 2003, Bell et al. 2004,
Cross et al. 2004, Conselice et al. 2005, Ferreras et al. 2005,
Thomas et al. 2005, Treu et al. 2005).
The detailed indices analysis of Thomas et al. (2005) shows that the more
massive early-type galaxies are generally dominated by old stellar 
populations, enriched in the $\alpha$ elements, while intermediate-age stars
are found in early-type galaxies less massive than 10$^{10}$ M$_{\odot}$ and
with low [$\alpha$/Fe] ratios (cf. also Nelan et al. 2005). The combination
of old stellar ages and high abundance ratios typical of massive early-type galaxies
is suggestive of short formation timescales (i.e. 1~Gyr, Thomas, Greggio \& Bender 1999), 
during which star formation occurred either with a flattened IMF 
(Thomas 1999), or through multiple bursts characterized by an enhanced 
star-formation activity. On the other hand,
the spheroidal morphology and hot dynamics observed in these systems require
major mergers, which result in a more extended dynamical history. The discrepancy
in the star-formation and dynamical timescales may be reconciled through the 
{\it progenitor bias} (van Dokkum \& Franx 2001), whereby about 50$\%$ of the present-day
early-type galaxies were transformed from star-forming galaxies at $z <$ 1, with
the latters possibly having a constant star formation rate prior to their morphological 
transformation. The question, at this point, is whether the scaling relations 
derived for early-type galaxies at $z <$ 1 apply also to early-type galaxies
at intermediate redshift. The high angular resolution
of the Advanced Camera for Surveys (ACS) onboard HST allows for the very first time
to measure the structural parameters of early-type galaxies in the UDF/GRAPES surveys
with redshifts between 0.5 and 1.3. In this paper, we will discuss their star formation
histories as derived from their GRAPES (GRism ACS Program for Extragalactic Science,
Pirzkal et al. 2004) spectra and their possible assembling histories as deduced from
their isophotal structure in the UDF images (a concordance cosmology,
$\Lambda$CDM, $\Omega_m=0.3$, $H_0=70$~km~s$^{-1}$~Mpc$^{-1}$, is assumed hereafter).

The traditional approach to the formation of early-type galaxies is to study their
stellar populations via spectro-photometric observables (see e.g. Worthey 1994; 
Trager et al. 2000; Bernardi et al. 2003), derive the age and metallicity
distribution of their (unresolved) stellar content and finally reconstruct their 
star-formation history. This is the ultimate ingredient in distinguishing between 
different mechanisms for the formation of ellipticals, such as monolithic 
collapse (Eggen, Lynden-Bell \& Sandage 1962) and hierarchical merging 
(Kauffmann et al. 1993). At the same time, the study of the morphology of early-type
galaxies allows us to constrain their assembly history.
\par
The advent of wide-field imaging surveys performed from the ground and with HST 
(e.g. Williams et al. 1996, Giavalisco et al. 2004, Beckwith et al. 2005) has proved 
that galaxy structure
evolves with time, so that more distant galaxies are more peculiar than those in
the local Universe on which the Hubble classification was originally based
(Driver et al. 1995, Glazebrook et al. 1995, Abraham et al. 1996, Driver et al. 1998, 
Brichmann \& Ellis 2000, Conselice et al. 2005). Furthermore, the morphology-density 
relation (Dressler 1980, Dressler et al. 1997) confirms that galaxy morphology depends 
strongly on environment, with early-type galaxies preferentially living in
high density regions. Notice, here, that the morphological K-correction can affect
the morphological classification, depending on the fading of the surface
brightness with wavelength.  Windhorst et al. (2002, cf. also Colley et al. 1996) 
showed that early-type galaxies endure a significant decrease in their surface
brightness from red to mid-UV wavelengths which could lead to a different 
classification and makes them almost undetectable at
intermediate to high redshifts. Mid-type and star-forming galaxies appear in  
the mid-UV with a somewhat later type, while the majority of late-type and merging 
systems have a morphology little dependent on wavelength.
\par\noindent
Galaxy shapes can be quantified into several ``morphological parameters''
(e.g. surface brightness profile, bulge-to-disk ratio, ellipticity,
light concentration, asymmetry and clumpiness) which turn out to correlate with
the galaxy star-formation rate, stellar mass, central black hole mass and
merging history. For example, the light concentration of a galaxy varies
with its luminosity, stellar mass, size and the mass of the central
black hole (Caon et al. 1993, Graham et al. 1996, Bershady et al. 2000, Graham
et al. 2001, Conselice 2003). Therefore, it depends on the
past formation history of a galaxy. Asymmetry and clumpiness ``measure'' a
more recent epoch in the evolution of a galaxy, since asymmetry is due to the
presence of a merger and/or to tidal interactions, and clumpiness correlates with
the degree of on-going star formation (Conselice 2003). These correlations 
allow us to decipher galaxy morphologies into mechanisms of galaxy formation and into
time evolution of galaxy assembly. 
\par
Extensive imaging
from the ground and with HST has revealed that a large fraction of nearby elliptical
galaxies (Es) differ from a pure ellipse shape and do show substructures. For example,
the ellipticity is not always constant with distance from the galaxy center. 
Di Tullio (1978, 1979) found that isolated Es are, on average, characterized by a
decreasing ellipticity, while the mean ellipticity profile is either increasing or peaking
with radius in Es located in galaxy clusters and groups.  A change in the position 
angle of the galaxy
major axis is often observed and explained as a projection effect of a triaxial 
structure whose
axes ratios ({\it b/a} and {\it c/a}) vary with radius (Galletta 1980, Franx 1988). 
About 60$\%$
of Es have ``non-elliptical'' isophotes, equally separated into {\it disky} or {\it boxy}
(Bender et al. 1988, 1989). Disky Es are generally oblate rotators, while
boxy Es span a variety of kinematical
properties and are systematically bigger and brighter, suggesting that they formed
via mergers (Naab, Burkert \& Hernquist 1999). 
Almost half of the nearby Es, especially those in galaxy groups,
show dust lanes which are aligned with the major or minor axis and thus believed
to be byproducts of mergers (Lauer 1985, Sadler \& Gerhard
1985, Ebneter \& Balick 1985, van Dokkum \& Franx 1995, Tran et al. 2001, Lauer et al. 2005, 
Martel et al. 2004). Elliptical
galaxies surrounded by shells have also been observed and are thought to be the result
of a collision with a disk galaxy (Malin \& Carter 1983,
Quinn 1984, Michard \& Prugniel 2004). Finally, a blue core, possibly experiencing 
star formation,
has been detected in $\sim$30$\%$ of nearby Es (Abraham et al. 1999, Papovich et al. 2003,
Menanteau et al. 2001, Goto 2005) and nuclear stellar disks in $\sim$60$\%$ of
early-type galaxies (van den Bosch et al. 1994, Rest et al. 2001, Lauer
et al. 2005).
\par
Our knowledge of the detailed structure of early-type galaxies is mostly based on what 
has been observed so far in the local Universe. The purpose of this paper is to extend this
study to early-type galaxies at intermediate redshift, so to describe in detail their structure - redshift 
correlation and to complement this with an accurate study of their stellar populations. 

%%%%%%%%%%%%%%%%%%%%%%%%%%%%%%%%%%%%%%%%%%%%%%%%
\section{Observations}
The Hubble Ultra Deep Field (UDF) is a survey of a 3$'$.4 $\times$ 3$'$.4
field located in the Chandra Deep Field South, which was carried out 
with the Advanced Camera for Surveys (ACS) in four filters: F435W, F606W,
F775W and F850LP (cf. Beckwith et al. 2005 for further details).
We followed up these observations with ACS grism and slitless
spectroscopy, as part of the GRAPES project which was awarded 40
HST orbits during Cycle 12 (ID 9793, PI S. Malhotra). The grism observations 
were taken at different epochs and at four different orientations
in order to minimize the spectra contamination and overlapping from
nearby sources. We have been then able to extract
low resolution spectra (R $\simeq$ 100, in the range 5500 \AA\/
to 10500 \AA) for 5138 sources in the UDF field down to a limiting
magnitude of $m_{F850LP} \simeq$ 29.5 in AB system. Readers are referred to
Pirzkal et al. (2004) for a complete discussion of the data acquisition
and reduction and for a description of the final GRAPES catalogue.

%%%%%%%%%%%%%%%%%%%%%%%%%%%%%%%%%%%%%%%%%%%%%%%%
\section{The sample of early-type galaxies}
Our project aims at a combined analysis of the spectroscopic and
morphological properties of early-type galaxies in the UDF. We
thus selected from the GRAPES catalogue those sources with a high 
signal-to-noise ratio in their spectra, typically
SNR$\sim 10-20$ per pixel at $\lambda =8000$ \AA; such values guarantee
an accurate analysis of the observed spectral energy distributions (SEDs) in
terms of stellar ages and metallicities. We found that this
criterion imposes an apparent magnitude limit around $m_{F775W}$ = 24.0 in
AB system. We then computed the concentration (C) and asymmetry (A) indices 
for the spectroscopically selected sample, in order to select those 
objects which lie in the parameter space defined by ellipticals (cf. Conselice 2003). 
Specifically, we derived the concentration and asymmetry parameters using the
images taken in the F775W band, which roughly maps into a rest-frame 
$B$ band at the redshifts reported in Table 1.   
We adopted for the concentration
and asymmetry the same definitions and computing recipes
given by Conselice (2003). We finally selected as {\it bona fide} early-type galaxies
18 objects, for which C $\geq$ 3.2 and A $\leq$ 0.2 in agreement with
Conselice et al. (2005), and whose elliptical morphology was also
visually confirmed. They are shown in Figure 1 together with 
the mean loci for nearby early-, late-type galaxies and mergers 
derived by Bershady et al. (2000) and Conselice et al. (2000). 

\par\noindent
Table 1 shows the main properties of our final sample,
including the UDF/GOODS ID numbers. Magnitudes are given in the AB system -- taking 
the MAG$_-$BEST values from SExtractor (Bertin \& Arnouts 1996).
The optical magnitudes come from the UDF catalogue (Beckwith et al. 2005).
Three of the galaxies in our
sample (J033229.22-274707.6,
J033240.33-274957.0, J033244.09-274541.5) are not in the UDF
field of view although their spectra were detected in the GRAPES survey.
Their images were extracted from the GOODS data (Giavalisco et al. 2004). 
UDF4527 and J033229.22-274707.6 have been also detected by Chandra
in the soft X-rays, with a flux of 3$\times$10$^{-17}$ erg~s$^{-1}$cm$^{-2}$
and 8$\times$10$^{-17}$ erg~s$^{-1}$cm$^{-2}$ respectively (Alexander et al. 2003,
Giacconi et al. 2002). 

The last two columns of Table 1 give the spectroscopic redshifts derived
with the ACS grism and with the VLT (FORS2: Vanzella et al. 2005, VIMOS:
Le~F\`evre et al. 2004). The redshifts obtained from the ACS grism
have been measured by matching templates corresponding to a range of stellar 
populations to the low resolution GRAPES spectra. For those galaxies 
with an accurate VLT redshift, our
grism estimates agreed to $\Delta z/(1+z)\simlt 0.01$, significantly
better than any photometric redshift estimate (see e.g. Mobasher et al. 2004).
The {\it bona fide} early-type galaxies span the redshift range between
$z \simeq$ 0.5 and 1.3, with about 80$\%$ of the sample being distributed
between $z \simeq$ 0.5 and $z \simeq$ 0.8.
Notice here that 39$\%$ of the {\it bona fide} early-type galaxies are clustered
around $z \simeq$ 0.665. This is a well known spike, since it has been detected
in the K20 sample of ellipticals (Cimatti et al. 2002), in the Chandra
Deep Field South (Gilli et al. 2003) and in the VLT spectroscopic follow-up of
GOODS (Vanzella et al. 2005). These surveys indicate that the galaxies at
$z \simeq$ 0.67 constitute a loose structure rather than a cluster.

%%%%%%%%%%%%%%%%%%%%%%%%%%%%%%%%%%%%%%%%%%%%%%
%%%%%%%%%% Star Formation Histories %%%%%%%%%%
%%%%%%%%%%%%%%%%%%%%%%%%%%%%%%%%%%%%%%%%%%%%%%
\section{Star formation histories}
Our first approach in the study of the sample involves the star formation 
histories. In order to break the age-metallicity degeneracy, we combined
models of galactic chemical enrichment and the low resolution slitless 
spectroscopy from the GRAPES data. We took extra care in avoiding 
contamination from nearby sources, which -- in some cases -- required
the use of data corresponding to a single orientation.

In order to make a robust assessment of the ages and metallicities of the unresolved 
stellar populations, we decided to use two different models to describe the star 
formation histories, as described below. Briefly, the models depend on
a reduced set of parameters, which uniquely determine a star formation history (SFH),
i.e. a distribution of stellar ages and metallicities.
This SFH is used to combine simple stellar populations from the models of
Bruzual \& Charlot (2003) and obtain a spectral energy distribution (SED) assuming
a Salpeter IMF. Each choice
of SFH is compared with the data via a $\chi^2$ test. We explore a wide volume of 
parameter space in order to infer robust constraints on the ages and metallicities
of the stellar populations in these galaxies. 
This comparison requires a careful process of degrading the synthetic SED 
(resolution $R\sim 2000$) to the (variable) resolution of the GRAPES spectra.
Special care must be taken with respect to the change of the PSF with wavelength, 
which results in both an effective degradation of the spectral resolution as a 
function of wavelength, and a different net spectral resolution with respect to
the size of the galaxy. The latter requires the spectral resolution to be treated
as a free parameter, to be chosen by a maximum likelihood method. We found that 
this parameter is not degenerate with respect to those describing the star formation
history, and mostly results in a global shift of the likelihood. The two models 
chosen to describe the build-up of the stellar component are as follows:

\noindent
{\bf Model \#1 (EXP):} We take a simple exponentially decaying star formation
rate so that each history is uniquely parameterized by a formation epoch 
-- which can be described by a formation redshift ($z_F$), a star formation 
timescale ($\tau_\star$) and a metallicity ($[m/H]$) which is kept 
fixed at all times. 

\noindent
{\bf Model \#2 (CSP):} We follow a consistent chemical enrichment code 
as described in Ferreras \& Silk (2000). The model allows for gas infall 
and outflows and keeps the star formation efficiency as a free parameter. 
The metallicity evolves according
to these parameters, using the stellar yields from Thielemann, Nomoto \& Hashimoto 
(1996) for massive stars ($>10M_\odot$) and van den Hoek \& Groenewegen (1997) 
for intermediate mass stars. The free parameters are the star formation efficiency,
the fraction of gas ejected in outflows, the formation epoch, and the timescale
for the infall of gas.

Figure 2 shows the best fits and the observed SEDs\footnote{The spectra 
of J033229.22-274707.6 and UDF6027
could resemble those of dusty ULIRGS (cf. Moustakas et al. 2004). The CAS morphology
of these two galaxies is still consistent with them being ellipticals although
somewhat close to the transition edge between early-type galaxies and ULIRGs.
The low resolution of the ACS grism certainly helps in smoothing the spectra
and decreasing the Balmer break.}. The error 
bars represent the observations and the solid line correspond to the best fits 
for the CSP model. Figure 3 shows the ages and metallicities of 
the best SFHs for each galaxy. 
Average and RMS scatter for age and metallicity are shown as dots and error bars,
respectively. Notice the metallicity in the EXP models do not have an error bar 
as this model assumes a fixed value of metallicity for each SFH. The lines in the 
bottom panel give the
age of the Universe as a function of redshift (solid line) and the ages at
that redshift corresponding to a formation epoch of (dashed, from top to bottom) 
$z_F=\{5,3,2\}$. Analogously to the analysis
of the surface brightness distribution shown in the next sections, we find that
the stellar populations in these galaxies are dominated by old, passively evolving
stars, formed at redshifts $z_F\simgt 2$. Notice that the EXP model tends to 
underestimate the stellar ages. This model assumes a constant metallicity, 
which often forces older populations to have higher metallicities, which is
compensated by the age-metallicity degeneracy by shifting the average towards
younger ages. The more consistent model (CSP) generates a more realistic
distribution of metallicities and we should expect it to be closer to
the true populations. Nevertheless, the difference between these models
is rather small, and we can safely reject a significant star formation at $z\sim 1-1.5$
in {\em all} galaxies. These models also give galaxy mass estimates in the
range 9.5 $\leq$ log(M/M$_{\odot}$) $\leq$ 11.5.  

The star formation histories that give the best fit can be used to 
predict the evolution of the color-magnitude relation (CMR). Figure 4
shows the rest-frame $U-V$ vs. $M_V$ CMR of our sample. The filled dots
are simple translations of the observed photometry to restframe $U-V$
(i.e. a K correction) whereas the open dots evolve these galaxies to
zero redshift, in order to compare our galaxies with local samples.
Two characteristic error bars are shown. We plot Coma cluster galaxies 
from Bower, Lucey \& Ellis (1992) and the dashed lines represent the 
linear fit and scatter. The Figure shows that our sample of 
early-type galaxies is fairly consistent with the stellar populations found 
in local elliptical galaxies. Notice that the passive fading of the
brightest galaxies puts them at $z=0$ within the luminosity range of the
brightest local ellipticals. The available information also allows us to
explore a projection of the Fundamental Plane, namely the Kormendy relation
(Kormendy 1977). Figure 5 shows this relation (filled dots
with error bars) compared to Coma cluster galaxies (squares; 
J\o rgensen, Franx \& Kj\ae rgaard 1995),
a sample of distant early-type galaxies in the HDF (grey dots; Fasano et al. 1998),
the LBDS radio elliptical galaxies at $z\sim 1.5$ (open stars; Waddington et al. 2002)
and the early-type galaxies at $z \sim$ 2 studied by Daddi et al. (2005, open dots).
The translation from the observed photometry to $\mu_B$ involves a K correction
as well as the $(1+z)^4$ cosmological dimming.
The offset $\Delta\langle\mu_B\rangle\sim 1.5$~mag between our sample and Coma 
(as well as between Coma and the HDF sample and the LBDS radio galaxies) is
compatible with the expected fading of the stellar populations. 
Notice that both the GRAPES and the HDF sample lack $r_{\rm hl}\sim 10$~kpc galaxies,
most likely due to the limited volume sample in the HDF and UDF surveys
(the $r_{\rm hl}$ of the {\it bona fide} early-type galaxies was measured in
the F775W filter, corresponding to the restframe B band at $<z> \simeq$ 0.7).
No {\it bona fide} early-type galaxy at $z \simeq$ 1 is as compact as those
studied by Daddi et al. (2005) at $z \sim$ 2.

%%%%%%%%%%%%%%%%%%%%%%%%%%%%%%%%%%%%%%%%%%%%%%%%
\section{Isophotal shapes}
We used IRAF ELLIPSE routine (Jedrzejewski 1987) to measure 
the isophotal parameters of the {\it bona fide} early-type galaxies
in the F606W, F775W and F850LP bands. At a mean redshift of 0.7 (cf. Table 1),
these filters sample the restframe U, B and V bands. ELLIPSE allows
the isophote center, ellipticity and position angle to vary at each iteration.
Once a satisfactory fit is found, it computes the sin and cos 3$\theta$ and
4$\theta$ terms, which describe the deviations of the isophote from pure
ellipse by means of the following Fourier expansion:
\par\noindent
$\Delta$$r(\theta)/r(\theta)$ = {\it (a$_3$/a)}cos(3$\theta$) $+$ {\it (a$_4$/a)}cos(4$\theta$)
$+$ {\it (b$_3$/b)}sin(3$\theta$) $+$ {\it (b$_4$/b)}sin(4$\theta$)
\par\noindent
where $\theta$ is the position angle, {\it r} the distance from the galaxy center
and {\it a} and {\it b} the semi-major and semi-minor axes.
The {\it a$_4$/a} and {\it a$_3$/a} parameters, where {\it a} is the semi-major
axis of the isophote, measure the deviations of the isophote from a pure ellipse.
In particular, {\it a$_4$/a} measures the deviations symmetrical with respect to the
galaxy center (i.e. found along the isophote every 90$^o$), while {\it a$_3$/a}
the non-symmetrical deviations occurring at every 120$^o$.
\par\noindent
We ran ELLIPSE
leaving free the position of the center, the ellipticity (defined as 
1 - {\it b/a}, where {\it a} and {\it b} are the projected major and
minor axes of a galaxy) and the position angle of the major axis and assuming
a logarithmic step along the semi-major axis. The results are shown for
each individual galaxy in Figures 11a - 11r (of which Figures 11b - 11r are
available in the electronic edition of the Journal), where the surface brightness,
ellipticity, position angle (PA) of the major axis, the {\it a$_3$/a} and 
{\it a$_4$/a} parameters are plotted for each filter as a function of distance from the 
galaxy center. These radial profiles extend to about 1$''$.2 from the galaxy
center in order to avoid overlap with nearby objects. 
\par
The ellipticity of the {\it bona fide} early-type galaxies show a variety of 
radial dependencies: increasing outward (i.e. J033229.22-274707.6, UDF9264,
UDF2322, UDF8, UDF153, J033240.33-274957.0, J033244.09-274541.5), decreasing
outward (UDF3677), peaked (i.e. UDF8316, UDF6747, UDF68, UDF4587, UDF4527, 
UDF2107, UDF5177), nearly flat (i.e. UDF2387, UDF1960, UDF6027). In the case
of UDF1960, UDF4587 and UDF5177, the radial profile of the ellipicity is
strongly perturbed by more compact (and usually bluer) sources projected onto
the main body of these galaxies. These features are somewhat dependent
on wavelength, since the ellipticity profiles appear to be smoother and to
extend to lower values in the restframe V band (i.e. the F850LP 
filter). Di Tullio (1978, 1979) analysed a sample of 75 nearby ellipticals 
and found that the radial variation of a galaxy ellipticity is different
for different environments. Indeed, ellipticals with increasing ellipticity
are found in galaxy clusters and groups, and systems with
peaked ellipticity also in galaxy pairs. Ellipticals with constant or
decreasing ellipticity are preferentially isolated, although are also 
observed in galaxy clusters, groups and pairs. Di Tullio (1978, 1979) suggested
that an increasing ellipticity with radius could be a signature of
merger and/or tidal interactions with a companion galaxy.
\par\noindent
The position angle of the {\it bona fide} early-type galaxies is seen to
generally vary with radius. The galaxies UDF1960, UDF3677, UDF68 and UDF8
show significant troughs in their PA profiles, which are likely due to
the presence of dust features and compact sources projected onto the main body 
of these galaxies (cf. Sect. 7). 
%The isophote twist in the PA profiles 
%of Figures 17 -- 34 is $\geq$ 10$^o$,
%independent of wavelength. 
Galletta (1980) explained the radial PA variations 
as the effect of the projection of a triaxial structure with
the axes ratios ({\it b/a} and {\it c/a}) varying with radius, onto the plane 
of the sky. Di Tullio (1979, 1980) proposed that PA variations larger than
10$^o$ could be relics of past mergers.
\par\noindent
Isophotes can locally vary from a simple ellipse shape: if this variation
occurs every 90$^o$ along the isophote and from the galaxy center, it 
indicates that the isophote has either a disky or a boxy shape. This
variation is quantitatively measured with the {\it a$_4$/a} parameter;
disky isophotes have {\it a$_4$/a} $>$ 0, boxy isophotes are characterized
by negative values of {\it a$_4$/a}, while elliptical isophotes have 
{\it a$_4$/a} = 0 (Lauer 1985, Bender et al. 1988, 1989).
In our sample of {\it bona fide} early-type galaxies,
the radial variation of the {\it a$_4$/a} parameter is quite dependent
on wavelength, except for the galaxies UDF1960, UDF3677, UDF8316, UDF6747,
UDF68, UDF2322, UDF6027 and UDF4527. The early-type galaxies which are
clearly disky ({\it a$_4$/a} $>>$ 0) at any wavelength are: UDF2387, UDF6747,
UDF4527, UDF2107 and UDF5177, while the only clear boxy galaxy
is UDF3677. More elliptical galaxies (with no significant variations at
any wavelength) are J033229.22-274707.6, UDF1960, UDF8316, UDF2322,
UDF6027, J033240.33-274957.0. The galaxies mentioned above may show troughs
or sharp peaks in the radial profile of their {\it a$_4$/a} parameter, 
which can be
produced by dust features and compact sources projected onto the main body of 
these galaxies. UDF153 is the only galaxy  that becomes boxy only in the 
restframe B and V bands (i.e. F775W and F850LP filters). Other galaxies
feature a radial change from disky to boxy (and vice-versa),
%change from being disky (elliptical) in the central parts to 
%boxy (disky) in the outer parts, 
such as UDF9264, UDF68, UDF8, UDF4587
and J033244.09-274541.5. In particular, UDF8 changes from elliptical to
disky in the restframe U and B bands (i.e. F606W and F775W filters), while
it changes from boxy to disky in the restframe V band (F850LP filter).
The radius at which this change occurs is in the area where the PSF
still dominates the light distribution of the galaxy (cf. Sect. 5).
\par\noindent
A second parameter, {\it a$_3$/a}, measures the deviation of an isophote
from a perfect ellipse, every 120$^o$ along the isophote and from the
galaxy center. This variation is mostly induced by the presence of dust
features and clumps in the galaxy. Among the {\it bona fide}
early-type galaxies, we detect {\it a$_3$/a} $\ne$ 0 in UDF2387,
UDF1960, UDF3677, UDF8316, UDF6747, UDF8, UDF6027, UDF153, UDF4587, 
UDF4527, J033244.09-274541.5, UDF2107 and UDF5177. 
\par\noindent
The measured isophotal parameters exhibit radial gradients that
do not appear to significantly depend on the wavelength at which the
galaxies were imaged. Given the caveat on the effects of the morphological
K-correction (Windhorst et al. 2002), this confirms the robustness of the
criteria we used to select our sample of {\it bona fide} early-type galaxies.
\par\noindent
Bender et al. (1988, 1989) analysed a sample of  nearby
ellipticals and concluded that they equally split among boxy systems
(30$\%$), disky galaxies (30$\%$) and ellipticals with no significant
deviations (30$\%$). In particular, disky ellipticals have ellipticity
larger than 0.25 and behave as oblate rotators, while boxy systems
have ellipticities in the range between 0.15 and 0.45 and display a large
variety of kinematical properties. In addition, boxy ellipticals are on
average bigger with half-light radii larger than 5 kpc and brighter 
than disky systems in the B band, at radio wavelengths and in the X-rays.
Starting from the idea whereby ellipticals can originate from
the merger of two disk galaxies (Toomre \& Toomre 1972), N-body simulations
of disks collisions predict kinematic and photometric properties of the 
merger remnants very similar to those observed for elliptical galaxies, i.e. 
decoupled cores and isophotal shapes (cf. Barnes 1992, Hernquist 1992,
Naab, Burkert \& Hernquist 1999, Naab \& Burkert 2003). Khochfar \& Burkert
(2005) have shown that the morphology of the merging galaxies and any subsequent
episode of gas infall are critical in determining the shape of the merger
remnant. Boxy ellipticals can be reproduced via a merger of two equally-massive
disk galaxies or a major merger of two early-type galaxies. Disky ellipticals
are mainly formed through an unequal-mass merger of two disk galaxies or 
late gas infall. In the hierarchical framework equal-mass mergers are
expected to be much rarer than the unequal-mass, although the observational
selection seems to go in the opposite direction (i.e. the ratio of boxy to
disky galaxies at $z \simeq$ 0 is nearly one).

%%%%%%%%%%%%%%%%%%%%%%%%%%%%%%%%%%%%%%%%%%%%%%%%
\section{Average isophotal parameters}
Given that the {\it bona fide} early-type galaxies in the UDF/GRAPES
survey are at higher redshift than the nearby ellipticals studied
by Bender et al. (1988, 1989), it is interesting to check whether 
the isophotal parameters show any change with redshift. We therefore
computed for each galaxy the characteristic ellipticity and isophotal twist
$\Delta$PA, the mean {\it a$_3$/a} and {\it a$_4$/a} parameters
({\it $\langle a_3/a\rangle$} and {\it $\langle a_4/a\rangle$}) 
following the prescriptions of Bender et al.
(1988) and in the F850LP filter which for the observed
redshifts corresponds to the restframe V and R bands used by Bender et al.
Briefly, we averaged the {\it a$_3$/a} and {\it a$_4$/a} values
measured at R$_i \leq r \leq$ R$_o$ weighted by their errors
and by the counts in the isophote at each isophotal mean radius 
$r$. Here, R$_o$ is 1.5 times 
the half-light radius r$_{hl}$ derived from the $\mu_{F850LP}$ radial profile
and R$_i$ is the radius corresponding to 3 times the FWHM of the PSF
in the F850LP filter, chosen to avoid the central part of a galaxy
where the light distribution is perturbed by the PSF. While R$_i$ is 12 pixels, 
i.e. 0$''$.36, R$_o$ varies up to 1$''$.1. 
%The {\it a$_4$/a} values were also corrected by a factor equal
%to {\it (b/a)}$^{0.5}$. 
The isophotal twist was measured as a difference
in PA between R$_i$ and R$_o$, while the characteristic ellipticity was
taken equal to the ellipticity at the half-light radius in case of
increasing, decreasing or flat radial profiles, or equal
to the peak ellipticity in case of peaked radial profiles. The
final results are reported in Table 2, where the {\it $\langle a_3/a\rangle$} and 
{\it $\langle a_4/a\rangle$} values
have been multiplied by 100. Since its R$_o$ is comparable to R$_i$, we 
could not determine the isophotal parameters of J033240.33-274957.0 and UDF68.
\par\noindent
As pointed out by Odewahn et al. (1997), the S/N ratio in the images can affect
the measurement of the galaxy projected {\it a} and {\it b} axes and hence
its ellipticity, in the sense that galaxies with a lower S/N value are measured
to be rounder than real.
Therefore, to test the reliability of the isophotal parameters
derived for our sample, we measured the ellipticity, {\it $\langle a_3/a\rangle$} and
{\it $\langle a_4/a\rangle$} parameters for a number of simulated ellipticals.
The simulations were performed (cf. Appendix A) using a S\'ersic profile (S\'ersic 1968)
with index between 3 and 4, and assuming the S/N ratio and the integrated magnitude of the
observed data. The simulated images were then convolved with the ACS PSF in the
F775W filter. We were able to recover for the simulated images not convolved by
the PSF isophotal parameters very close to their input values, thus indicating that
the S/N ratio of the data does not significantly affect our measurements.
We also found that the ACS PSF underestimate the ellipticity by $\simeq$10$\%$,
overestimates {\it $\langle a_4/a\rangle$} by about 0.2 with respect to the null input
value and introduces a scatter of about 0.2 in {\it $\langle a_3/a\rangle$} 
around the input {\it a$_3$/a} = 0. used for the simulations.

The isophotal parameters of the {\it bona fide} early-type galaxies 
(black filled circles) are
plotted in Figure 6 as a function of the galaxy half-light radius
r$_{hl}$ (measured in the F850LP filter, corresponding to the restframe V band at 
$<z>\/ \simeq$ 0.7), 
total absolute magnitude in restframe B band M$_B$,
and compared with the sample of Bender et al. (1988, 1989)
represented with grey, filled circles. It has to be specified that the
M$_B$ values were derived from the best-fitting, model spectra
(cf. Sect. 4), also K-corrected. The magnitude range (hence stellar mass) of the nearby 
ellipticals is nicely sampled by the {\it bona fide} early-type galaxies, and this
makes the comparison of the two galaxy samples in their isophotal parameters even
more robust. We chose to correct the ellipticity and the {\it $\langle a_4/a\rangle$}
parameter for the PSF, i.e we increased ellipticities by 10$\%$ and decreased 
{\it $\langle a_4/a\rangle$} by 0.15. Overall, the {\it bona fide} early-type
galaxies confirm the trends traced by nearby ellipticals:
\par\noindent
%{\it i)} the isophotal twist increases at lower ellipticity, given the
%uncertainties in determining the major axis PA of a rounder galaxy.
%\par\noindent
{\it i)} boxy ellipticals tend to be bigger (as shown in Figure 6a).
\par\noindent
{\it ii)} Rounder ellipticals show smaller deviations from pure
ellipses, while larger deviations are seen in
galaxies with ellipticity $\geq$ 0.25. Disky ellipticals have 0.15 $\leq$
ellipticity $\leq$ 0.6, while boxy ellipticals 0.15 $\leq$ 
ellipticity $\leq$ 0.45. The deviant galaxies in our sample are UDF1960
and UDF6747 (cf. Figure 6b). 
The ellipse fitting of UDF1960 is affected by the presence of a blue
clump (a probable lens candidate, cf. Sect. 7.4) at a distance of 0$''$.4 from the 
galaxy center, and by the disk-like substructure in UDF6747 (cf. Figure 10).
\par\noindent
{\it iii)} Boxy ellipticals tend to be brighter than disky systems (see Figure 6c).
\par\noindent
Although not shown in Figure 6, the isophotal twist increases at lower ellipticity
because of the uncertainties in determining the major axis PA of a rounder galaxy.
We can thus conclude that ellipticity and {\it $\langle a_4/a\rangle$} do not
significantly change with redshift up to $z \simeq$ 1.
About half of the {\it bona fide} early-type galaxies are characterized
by values of {\it $\langle a_3/a\rangle$} larger than the range spanned
by nearby ellipticals. Their large {\it $\langle a_3/a\rangle$} values are likely due
to the presence of dust features (e.g. UDF2387 and UDF3677, cf. Sect. 7.1) or
blue clumps close to the galaxy center (e.g. UDF4587 and UDF2107,
cf. Sect. 7.3; UDF1960 with an overlapping lens candidate is a special case, cf. Sect. 7.4).
When dust features, disk-like substructures and blue clumps are masked out, the
isophote fitting gives significantly smaller values of the mean {\it $\langle a_3/a\rangle$} 
parameter, which then falls in the range spanned by nearby ellipticals.

%%%%%%%%%%%%%%%%%%%%%%%%%%%%%%%%%%%%%%%%%%%%%%%%
\section{Substructures in the UDF early-type galaxies}
We used the outputs of the ELLIPSE routine to build model images of
the {\it bona fide} early-type galaxies, and subtracted them from
the observed frames in order to look for residual substructures.
In the cases where substructures were detected, we re-ran 
ELLIPSE masking out the substructures and adjusted
the mask every run until the extinction map of each galaxy reached a 
minimum value of $\simeq \pm$ 0.05 mag in the unmasked pixels. The
extinction map was computed from the ratio of the observed image to the
ELLIPSE model image with the formula: 2.5 $\times$ Log(Observed/Model). 
We applied this procedure to the
images obtained in the F606W and F775W filters, which are characterized
by a better S/N ratio and PSF with respect to the F435W and
F850LP filters.
\par
In this way, we could detect dust features, disks and clumps in a
significant fraction of our sample. Six galaxies do not show either
a disky substructure nor dust features:
J033229.22-274707.6, J033240.33-274957.0, J033244.09-274541.5,
UDF2322, UDF6027 and UDF8316. J033244.09-274541.5 is shown in Figure 9:
a close-up of the central region (1$''$.2 $\times$ 1$''$.2) in 
the F775W filter is in the left-hand panel in units of counts/s, 
and the extinction map 
of the same region in units of magnitude (mean residual value $\simeq \pm$ 0.05
mag/pix) in the right-hand side panel. The (V-I) color maps of the
six featureless galaxies together with the remainder of our sample
are shown in Figures 7 and 8. 
The (V-I) color maps were obtained using Voronoi tessellations,
so that pixels were binned in order to reach a chosen and constant
S/N ratio per bin (cf. Cappellari \& Copin, 2003). In our case, the
targeted S/N ratio corresponds to 0.05 mag the color uncertainty.

\vskip 0.5truecm
\par\noindent
\subsection{Dust features}
Dust features have been resolved in UDF2387 and UDF3677 (see Figure 9).
They appear as white stripes in the extinction maps (right-hand side panels
of both Figures). Here, grey indicates a residual of about 0.05 mag and white 
an extinction A$_{F775W}$ ($\simeq$ restframe A$_B$) up to $\simeq$ 0.25 mag
in UDF2385 and 0.18 mag in UDF3677. These values are very
similar to what found for dusty nearby ellipticals at $z \simeq$ 0 by
Sadler \& Gerhard (1985).

\subsection{Disks}
UDF4527 and UDF6747 are characterized by a disk-like structure
that is aligned with the galaxy major axis and has a full extension
of about 1$''$ (corresponding nearly to the half-light radius
of each galaxy). These
features are shown in Figure 10,  where they turn out to be in 
emission (black pixels) 
by $\sim$ 0.10 mag/pix
in UDF4527 and $\sim$ 0.25 mag/pix in UDF6747.
In these images, light grey levels indicate residuals of about $\pm$
0.05 mag. In the (V-I) [restframe (B-V)] color map of UDF4527 and 
UDF6747, the disk turns out to be as red as the galaxy and the two should 
thus be coeval. Such a similiarity is known to be common among
S0 galaxies, because of either small age differences (Peletier \& Balcells 1996)
or a larger amount of dust in the disk (Michard \& Poulain 2000).

\subsection{Blue clumps}
As seen in Figures 7 and 8, 
a large fraction of {\it bona fide} early-type
galaxies shows compact, blue clumps at different distances (up to
2$''$) from the galaxy 
center. UDF6747, in particular, is surrounded by a cluster of blue clumps
at one edge of its major axis (cf. Figure 8). Figure 10  
highlights this feature in UDF153.
We could measure the integrated magnitude of these clumps in the four
filters of the UDF after having subtracted the ELLIPSE model image from
the observed frame of each galaxy. It turns out that these clumps are
very faint, with magnitudes between 30.9 and 26.1, and contribute to
the total flux of the host galaxy at a $\leq$ 4$\%$ level.  
Hence, these clumps do not
contribute significantly to the grism spectra of the {\it bona fide}
early-type galaxies. What is the nature of the blue clumps? The simplest
explanation would be that they are sources at high redshift, as in the
case of UDF1960, where the blue clump (cf. Figure 7) is a lens candidate
(Fassnacht et al. 2004). On the other hand,
given the number of objects detected in the UDF field and the size of
the galaxies in our sample, we have estimated a probability of about
30$\%$ for objects to overlap. A more intriguing possibility is that
the most compact of these clumps are physically associated with the {\it bona fide}
early-type galaxies, and, as such, may be HII regions or young star clusters 
or satellite dwarf galaxies. To test this hypothesis, we ran the
Hyperz code (Bolzonella et al. 2000) on the set of magnitudes measured for each clump,
assuming that the clump is at the same redshift of the galaxy in which
it has been observed. The reduced $\chi^2$ of the fits indicates that
some of these clumps could be consistent with the above assumption,
and they could be young (few 10$^6$ -
few 10$^8$ yrs) systems in the range -12 $\leq$ M$_V \leq$ -16. 
Though speculative, these results
are in good agreement with the study of O'Dea et al. (2001) who found
four star-forming regions in the double-double radio galaxy 3C 236, and of
Yang et al. (2004) who detected
blue compact sources in five E$+$A galaxies at $z \simeq$ 0.1.

\subsection{Peculiar features}
UDF68, UDF1960, UDF2107 and UDF5177 display in their
extinction map some arc-like features in emission. An example is given
in Figure 10 for UDF5177. With the data in hand, it is not possible
to establish whether these features are real substructures or, for
example, lensed sources at high redshifts overlapping with the UDF 
early-type galaxies. Fassnacht et al. (2004) have identified only one lens
candidate among the {\it bona fide } early-type galaxies, which lies nearby 
UDF1960. We suggest that UDF68 has also lensed sources projected onto its main
body, given that the morphology of its (V-I) map is quite similar to UDF1960. 
The arc feature in UDF5177 is apparently offset from the center of the light
distribution of the galaxy and could be interpreted as the remnant of an accreted
dwarf galaxy which went tidally disrupted.

\subsection{Blue cores}
Two galaxies in our sample exhibit a blue core: UDF5177
and UDF6027 (cf. Figure 8). About 30$\%$ of nearby ellipticals have a blue
core, which is usually believed to have experienced recent star formation
(Abraham et al. 1999, Papovich et al. 2003, Menanteau et al. 2001, Goto 2005). 
Menanteau et al. (2004) found a blue-core elliptical at 
$z =$ 0.624 which seems to host an AGN. In our sample, UDF5177 and
UDF6027 are at $z =$ 0.50 and 1.32, respectively.

%%%%%%%%%%%%%%%%%%%%%%%%%%%%%%%%%%%%%%%%%%%%%%
%%%%%%%%%%%%%%%   Conclusions   %%%%%%%%%%%%%%
%%%%%%%%%%%%%%%%%%%%%%%%%%%%%%%%%%%%%%%%%%%%%%

%%%%%%%%%%%%%%%%%%%%%%%%%%%%%%%%%%%%%%%%%%%%%%%%
\section{Conclusions}
According to the current paradigm of galaxy formation, ellipticals form
through the merger of galaxies. As shown by Kauffmann \& Charlot (1998)
this predicts a fairly extended assembly history, which is difficult to
reconcile with the old stellar age and the $\alpha$-enhanced chemistry of these
galaxies. Because of their merger history, one would expect the isophotal 
morphology of early-type galaxies to be more distorted at high redshift compared to
nearby ellipticals. In this respect, the UDF survey is an unique dataset which
allows, for the very first time, to measure accurate isophotal parameters for
18 {\it bona fide} early-type galaxies between $z \simeq$ 0.5 and 1.3. At
the same time, the GRAPES survey provides for these same objects high S/N grism spectra,
which are used to determine the their stellar age and formation redshifts.
The population of $z\sim1$ early-type galaxies in the UDF is clearly old.
The analysis of their spectra indicates that their stellar populations
are rather homogenous in age and metallicity and formed at redshifts $z_F\sim 2-5$. 
Evolving them passively, they become indistinguishable
from the ellipticals observed in the Coma cluster [cf. also
Daddi et al. (2005) for a sample of UDF early-type galaxies at $z \sim$ 2.]
As discussed by Hamilton (1985), spectroscopically-selected early-type
galaxies are generally seen long after (i.e. $>>$ 1 Gyr) the last major merger.
After such a merger, these objects quickly settle back to their quasi steady-state, and
will then look for a much longer time like non-evolving ellipticals: it is during this
phase that we select and observe them now (cf. also van Dokkum \& Franx 2001).
%Therefore, ellipticals
%seem to have formed at high redshift and evolved passively since, 
%as also discussed by Daddi et al. (2005) for a sample of UDF early-type
%galaxies at $z \sim$ 2.
\par
The $z\sim1$ early-type galaxies in the UDF are morphologically
very similar to nearby ellipticals. They separate into disky galaxies
($\sim$ 44$\%$ with {\it $\langle a_4/a\rangle$} $>$ 0.5), 
boxy galaxies ($\sim$ 31$\%$ with {\it $\langle a_4/a\rangle$} $<$ -0.5) 
and systems with no significant deviation from a pure ellipse ($\sim$ 25$\%$
with -0.5 $\leq$ {\it $\langle a_4/a\rangle$} $\leq$ 0.5). Given the 
small field of view of the UDF and therefore the small sample, these fractions
can be taken as in good agreement with the results of Bender et al.
(1988, 1989) who found $\sim$30$\%$ of the nearby ellipticals to be disky, 
$\sim$30$\%$ boxy and $\sim$30$\%$ with no significant deviations. 
The slight overabundance of disky {\it vs} boxy ellipticals in our sample
could be explained by their environment. Indeed, VLT spectroscopic follow-up has 
detected three large-scale structures in the UDF (at $z \simeq$ 0.67,
0.73, 1.10 and 1.61, Vanzella et al. 2005, see also Cimatti et al. 2002 and
Gilli et al. 2003) which make the UDF an
over-dense region at any of these three redshifts. In over-dense regions,
the N-body simulations of Khochfar \& Burkert (2005) predict a higher number of disky 
{\it vs} boxy ellipticals.  The isophotal structure
of the {\it bona fide} early-type galaxies obeys the correlations already
observed among nearby ellipticals: {\it i)} disky ellipticals have generally
higher characteristic ellipticities; {\it ii)} boxy ellipticals have larger
half-light radii and are brighter in the B band. In this respect, 
the isophotal shape of ellipticals does not significantly change 
between $z =$ 0 and 1. This suggests that these galaxies either do not merge 
significantly in their evolution towards $z =$ 0, or mergers at later redshifts 
do not vary the percentages of disky and boxy ellipticals. 
In this respect, Schade et al. (1999) suggested that 
merging is not required since $z =$ 1 to produce the present-day space density
of ellipticals.
\par\noindent
The {\it $\langle a_3/a\rangle$} parameter appears to be
larger than measured by Bender et al. for nearly
one third of the {\it bona fide} early-type galaxies. This deviation is likely due
to the presence of dust features (in $\sim$ 10$\%$ of the sample, a percentage a factor of 4
smaller than observed at $z$ = 0) and most notably of clumps close to the galaxy
center. Blue (in the (V-I) color) clumps have been detected in about 50$\%$ 
of the {\it bona fide} early-type galaxies, at different distances from the
center (up to 2$''$, in the galaxy outskirts). From their 
photometry and under the assumption that they
are at the same redshift as their host galaxies, we have derived ages 
(from few 10$^6$ to few 10$^8$ yrs) and luminosities ( -12 $\leq$ M$_V \leq$ -16)
consistent with them being either young star clusters or dwarf irregular galaxies. 
They appear very similar to the blue compact sources observed in 
five E$+$A galaxies at $z \simeq$ 0.1 by
Yang et al. (2004), who suggested these sources to be young star clusters.
On the other hand, the hypothesis of being dwarf galaxies would put these blue
clumps in the right mass range to explain the small mass increase undergone by 
ellipticals more massive than 10$^{11.5}$ M$_{\odot}$ (i.e. these galaxies
seem to have grown less than 1$\%$ of their mass since $z \simeq$ 1
according to the results of Treu et al. 2005, or less than 5$\%$ as discussed
by Schade et al. 1999). Gas accretion is also
invoked to explain the presence of four star-forming regions in the
double-double radio galaxy 3C 236; their age difference would compare
with the timescale of the nuclear activity in this galaxy and hence
would set time constraints on the infall of gas experienced by
3C 236 (O'Dea et al. 2001).
\par
To add more speculation to the picture, since the blue clumps in our sample
galaxies are distributed at 
different distances from the galaxy center, we could imagine that they have been 
accreted and may possibly sink to the galaxy center, forming a blue core as observed
in UDF5177 at $z \simeq$ 0.5. We might then be facing the build-up of early-type 
galaxies at $z \simeq$ 1, similarly to what observed by Chen et al. (2003), Bell et al.
(2004), Cross et al. (2004), Conselice et al. (2005) and Ferreras et al. (2005).
Clearly, a dedicated spectroscopic campaign would 
be needed to unveil the true nature of these blue clumps.
\par
In the sample of {\it bona fide} early-type galaxies, UDF153 is particularly
remarkable, since it is surrounded by a faint and incomplete shell. As shown 
by Quinn (1984), such a shell is the relic of a relatively recent merger with 
a smaller {\it disk galaxy} and can be observed for about 1 Gyr. The high phase-space
density of a disk galaxy is indeed required to produce such a spatially confined 
substructure as the observed shells are.
The GRAPES spectrum
of UDF153 indicates a redshift of 0.98, suggesting that the merger possibly
occurred at $z \geq$ 1. This would also imply that some disk galaxies were already 
in place at $z \geq$ 1, despite observations which show that the fraction of
regular disk galaxies decreases at $z >$ 1 (cf. Abraham et al. 1996, Conselice et al. 
2003).
\par
N-body simulations (cf. Khochfar \& Burkert 2005, Jesseit et al. 2005) indicate
that the boxyness/diskyness shape of an elliptical galaxy is a long-lived signature
of a merger if no further disturbances (i.e. late gas infall and/or subsequent
mergers) occur. Indeed, the stellar orbits which forge the {\it a$_4$/a} parameter
are stable and in equilibrium. Therefore, if the system stays 
unperturbed after the major merger which formed it, the ``lifetime'' of its
isophotal shape is comparable with the age of its stellar population. Moreover,
the isophotal shape of the merger remnant depends on the mass ratio of the merger
and on the morphology (i.e. stellar orbits) of the merging galaxies.
Given that the fractions of disky and boxy systems among the {\it bona fide} early-type 
galaxies is similar to the one observed at $z =$ 0, and modulo the small size of 
their sample, the {\it bona fide} early-type galaxies can be used to trace the mass 
ratio of mergers as a function of redshift and the lifetime of the different isophotal
shapes taken by ellipticals in N-body simulations like those of Khochfar \& Burkert.
\par
A last comment is devoted to the masses of the {\it bona fide} early-type galaxies. 
Our spectroscopic analysis indicates masses between 10$^{9.5}$ and 10$^{11.5}$
M$_{\odot}$ in stars, which likely formed at about $z \simeq$ 3. At this redshift,
such masses are consistent with Lyman-break galaxies (LBG, Steidel et al. 1996,
Shapley et al. 2001, Papovich et al. 2001), SCUBA galaxies (Bertoldi et al. 2000,
Smail et al. 2002, Genzel et al. 2003, Webb et al. 2003, Chapman et al. 2003, 2004)
and bright radio galaxies (van Breugel et al. 1998, Dey et al. 1997). These galaxies
are all observed undergoing 
intense star formation with a rate varying from $\sim$ 90 M$_{\odot}$yr$^{-1}$ 
(in LBGs, Shapley et al. 2001) up to 500 M$_{\odot}$yr$^{-1}$ (in SCUBA sources,
Genzel et al. 2003) and 1000 M$_{\odot}$yr$^{-1}$ (in high-redshift radio galaxies,
Dey et al. 1997). Such a star-formation rate is able to enrich the galactic ISM up to a 
metallicity of about
0.4~Z$_{\odot}$ (Pettini et al. 2001), a value very similar to what found in this
paper for the {\it bona fide} early-type galaxies. Such starbursts are also likely  
to produce the $\alpha$-element enhancement typical of elliptical galaxies
(Trager et al. 2000).
These similarities in mass and metallicity add further evidence to the evolution
of vigorously star-forming galaxies at high redshift into $z \simeq$ 1 ellipticals.
%Therefore, we speculate that the UDF {\it bona fide} early-type 
%galaxies at $z \simeq$ 0.7, like possibly the UDF early-type galaxies at $z \sim$ 2
%(Daddi et al. 2005), may be the {\it observable} descendants of $z =$ 3.

\acknowledgments
We would like to thank an anonymous referee for useful comments that improved the paper.
We also thank A. Burkert, F.C. van den Bosch and C. Porciani
for valuable discussions. ED acknoledges support from NASA through the
Spitzer Fellowship Program, under award 1268429.
The imaging and spectroscopy data are based on observations with the
NASA/ESA {\it Hubble Space Telescope}, obtained at the Space Telescope
Science Institute, which is operated by AURA Inc., under NASA contract
NAS 5-26555. This work was supported by grant GO-09793.01-A, GO-09793.03-A
and GO-09793.08-A from the Space Telescope Science Institute. This
project has made use of the aXe extraction software, produced by ST-ECF,
Garching, Germany.

\citeindexfalse

\clearpage
\begin{deluxetable}{lcccccccc}
\tablecolumns{9}
%\rotate
\tablewidth{0pc}
\tablecaption{The sample: UDF/GOODS IDs, coordinates and magnitudes (AB)}
\tablehead{
\colhead{ID} & \colhead{RA(J2000)} & \colhead{Dec(J2000)} & \colhead{F435W} &
\colhead{F606W} & \colhead{F775W} & \colhead{F850LP} &  \colhead{$z_{GRISM}$} 
& \colhead{$z_{VLT}$}}
\startdata
J033229.22$-$274707.6 & 03:32:29.2 & $-$27:47:07.6 & 22.22 & 22.86 & 20.97 &
20.99 & 0.67 & 0.67\\
UDF 2387 & 03:32:35.8 & $-$27:47:58.8 & 24.52 & 22.13 & 20.74 & 20.29 & 0.66 & 0.66\\
UDF 1960 & 03:32:36.0 & $-$27:48:11.9 & 23.95 & 22.55 & 21.59 & 21.29 & 0.60 & 0.60\\
UDF 9264 & 03:32:37.2 & $-$27:46:08.1 & 22.78 & 23.46 & 21.96 & 20.89 & 1.07 & 1.10 \\
UDF 3677 & 03:32:37.3 & $-$27:47:29.3 & 24.21 & 21.74 & 20.30 & 19.81 & 0.65 & 0.67\\
UDF 8316 & 03:32:38.4 & $-$27:46:31.9 & 22.33 & 23.48 & 22.21 & 21.78 & 0.62 & \\
UDF 6747 & 03:32:38.8 & $-$27:46:48.9 & 25.22 & 23.03 & 21.73 & 21.28 & 0.54 & 0.62\\
UDF 68 & 03:32:38.8 & $-$27:49:28.5 & 26.83 & 24.49 & 22.98 & 22.36 & 0.82 & \\
UDF 2322 & 03:32:39.2 & $-$27:47:58.4 & 24.93 & 22.49 & 21.07 & 20.62 & 0.66 &\\
UDF 8 & 03:32:39.5 & $-$27:49:28.3 & 24.85 & 22.68 & 21.39 & 20.96 & 0.65 &\\
UDF 6027 & 03:32:39.6 & $-$27:47:09.1 & 26.19 & 24.96 & 23.71 & 22.79 & 1.15 & 1.32\\
UDF 153 & 03:32:39.6 & $-$27:49:09.6 & 25.15 & 22.95 & 21.51 & 20.63 & 0.98 & 0.98 \\
J033240.33$-$274957.0 & 03:32:40.3 & $-$27:49:57.0 & & 23.51 & 22.23 & 21.88 & 0.66 &\\
UDF 4587 & 03:32:40.7 & $-$27:47:31.0 & 23.99 & 22.96 & 21.67 & 21.22 & 0.67 & 0.67\\
UDF 4527 & 03:32:41.4 & $-$27:47:17.2 & 24.42 & 22.41 & 21.06 & 20.61 & 0.62 &\\
J033244.09$-$274541.5 & 03:32:44.1 & $-$27:45:41.5 & 23.11 & 21.12 & 20.09 & 19.75 & 0.49 &\\
UDF 2107 & 03:32:45.8 & $-$27:48:12.9 & 24.42 & 22.77 & 21.76 & 21.38 & 0.62 & 0.53\\
UDF 5177 & 03:32:48.5 & $-$27:47:19.6 & 24.74 & 23.33 & 22.43 & 22.08 & 0.50 &\\
\enddata
\end{deluxetable}

\begin{deluxetable}{lrrrr}
\tablecolumns{5}
\tablewidth{0pc}
\tablecaption{The isophotal parameters of the {\it bona fide} early-type galaxies. These
have not been corrected for the errors derived in the Appendix.}
\tablehead{
\colhead{ID} & \colhead{Ellipticity} & \colhead{{\it $\langle a_3/a\rangle$} $\times$ 100} &
\colhead{{\it $\langle a_4/a\rangle$} $\times$ 100} & \colhead{$\Delta$PA}}
\startdata
J033229.22-274707.6 & 0.14 & -0.48 &  0.23 & 1.39\\
UDF 2387            & 0.14 & -1.13 &  0.83 & 15.17\\
UDF 1960            & 0.20 &  5.75 &  3.59 & 0.23\\
UDF 9264            & 0.26 & -0.58 & -0.59 & 0.66\\
UDF 3677            & 0.08 & -0.68 & -0.65 & 5.60\\
UDF 8316            & 0.17 &  0.23 & -0.15 & 0.41\\
UDF 6747            & 0.20 &  1.04 &  3.61 & 3.26\\
UDF 68              &  &   &   & \\
UDF 2322            & 0.13 &  0.65 &  0.28 & 1.96\\
UDF 8               & 0.29 & -0.78 &  0.47 & 0.64\\
UDF 6027            & 0.06 &  0.66 & -0.77 & 68.29\\
UDF 153             & 0.26 & -0.10 & -1.98 & 15.90\\
J033240.33-274957.0 &  &  &  & \\
UDF 4587            & 0.41 & -1.93 & 0.46 & 0.50\\
UDF 4527            & 0.36 & -0.11 & 1.90 & 6.69\\
J033244.09-274541.5 & 0.30 & -0.48 & 0.60 & 4.76\\
UDF 2107            & 0.22 & -0.50 & 1.53 & 0.07\\
UDF 5177            & 0.19 & -0.85 & 2.05 & 37.28\\
\enddata
\end{deluxetable}

\clearpage
\appendix
\section{Errors on the measured isophotal parameters}
When comparing the isophotal parameters of galaxies at different redshift
and observed with different instruments, it becomes important to
qualitatively establish the systematics in the data. For this reason,
we simulated elliptical galaxies using a S\`ersic profile (S\`ersic 1968)
with an index
chosen between 3 and 4, an integrated magnitude and corresponding S/N
ratio taken from the range spanned by the data.
The input ellipticities and position angles were
chosen among those measured for the {\it bona fide} early-type galaxies
and the input {\it a$_3$/a} and {\it a$_4$/a} parameters were rigorously
set to 0. The simulated galaxies were then convolved by the ACS PSF in
the F775W filter; this PSF is the average image of stars in the UDF F775W
frame.
\par
We first ran ELLIPSE on the set of simulated galaxies convolved by the
PSF and measured the isophotal parameters with the method described
in Sect. 5. We then repeated the measurements on the simulated galaxies NOT
convolved by the PSF. The results are summarized in Figures 12
and 13.
When no PSF is applied, the output isophotal parameters agree extremely
well with their respective input values, thus indicating that the S/N ratio of 
the data does not significantly affect our measurements (but see also
Odewahn et al. 1997). The convolution with the PSF
introduces deviations from the input parameters so that:
\par\noindent
{\it i)} the output ellipticities are underestimated by about 10$\%$;
\par\noindent
{\it ii)} the output PA values are underestimated by about 3$\%$;
\par\noindent
{\it iii)} the output {\it a$_3$/a} parameter is scattered by about 0.2
around the input null value;
\par\noindent
{\it iv)} the output {\it a$_4$/a} parameter is also overestimated
by an amount which increases with the input ellipticity. For
$\epsilon <$ 0.5 (the range in Table 2) we derive a discrepancy
of $+$0.15 in {\it a$_4$/a}.
\par\noindent
These deviations do not show any correlation with the input magnitude
and S\`ersic index and represent the sistematic uncertainties on the
values reported in Table 2.

\clearpage
\begin{figure}
\begin{center}
\includegraphics[width=3.8in]{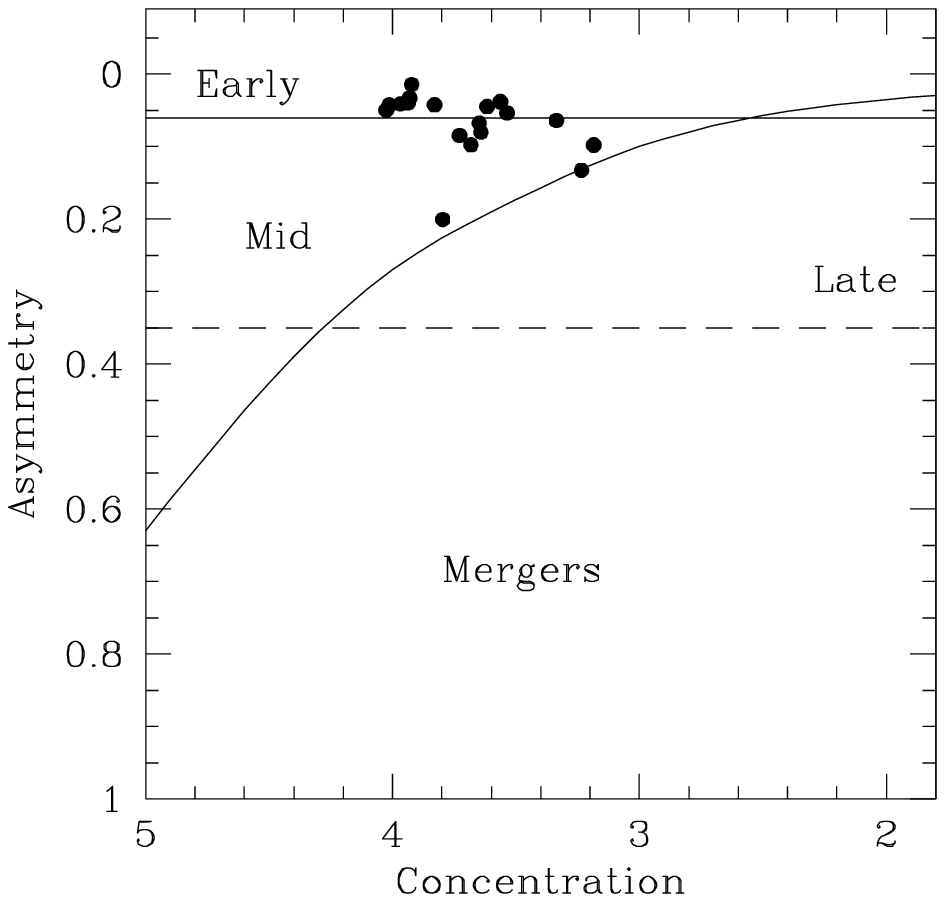}
\end{center}
FIG. 1.- Asymmetry and Concentration values for
the  {\it bona fide}
early-type galaxies. The distinction between
early-, late-type, mid galaxies and mergers is from
Conselice et al. (2005).
\label{fig:CAS}
\end{figure}

%\clearpage
\begin{figure}
\begin{center}
\includegraphics[width=3.8in]{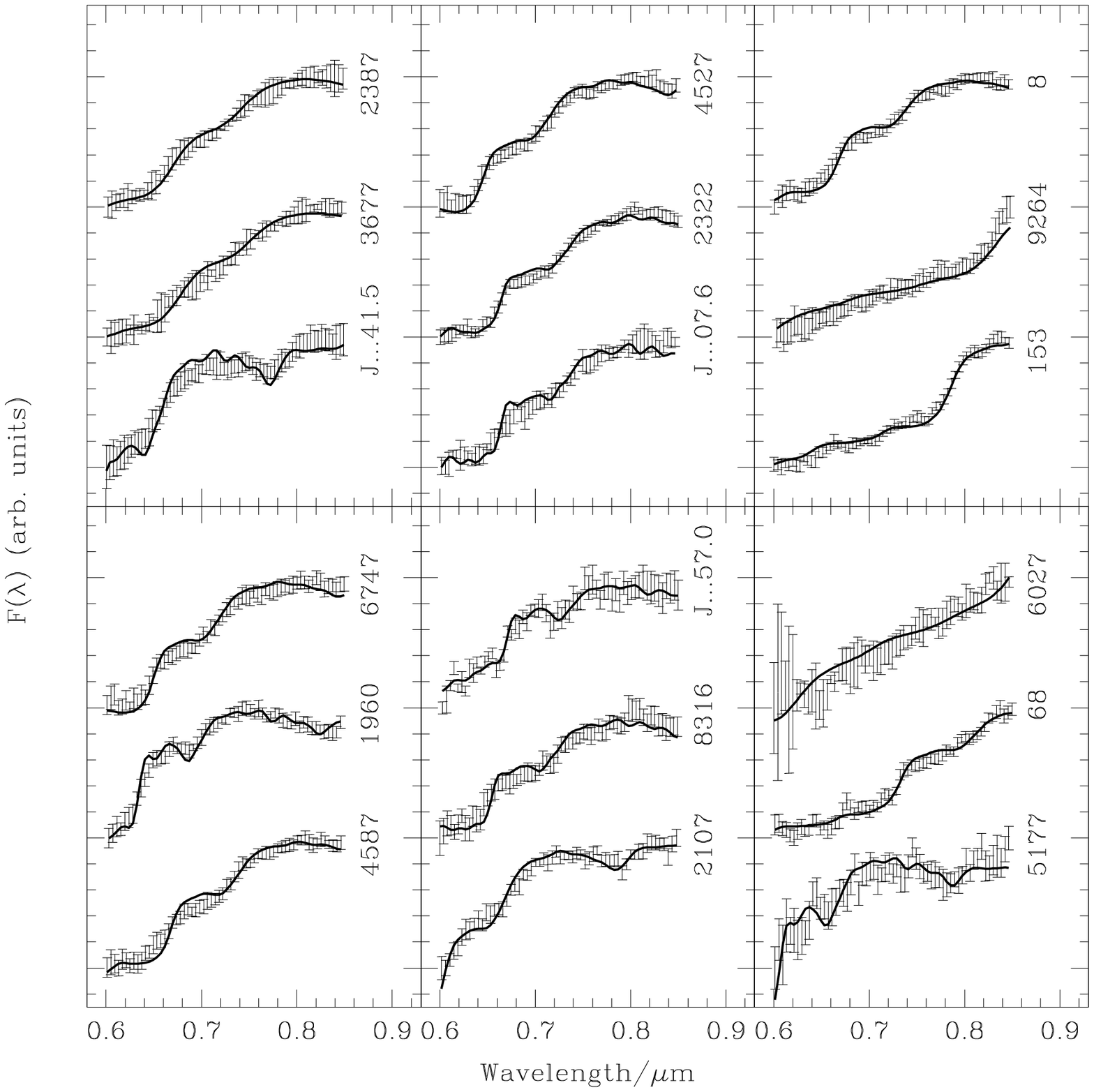}
\end{center}
FIG. 2.- Spectral energy distributions of the 18 galaxies in our sample (see
Table 1. Each one is labeled with the UDF numbers, except for those
outside of UDF, for which we give part of the GOODS identification. The error bars
are the observed ACS/G800L data and the lines are the best fits according to
the CSP models (see text for details).
\label{fig:seds}
\end{figure}

\clearpage
\begin{figure}
\begin{center}
\includegraphics[width=3.8in]{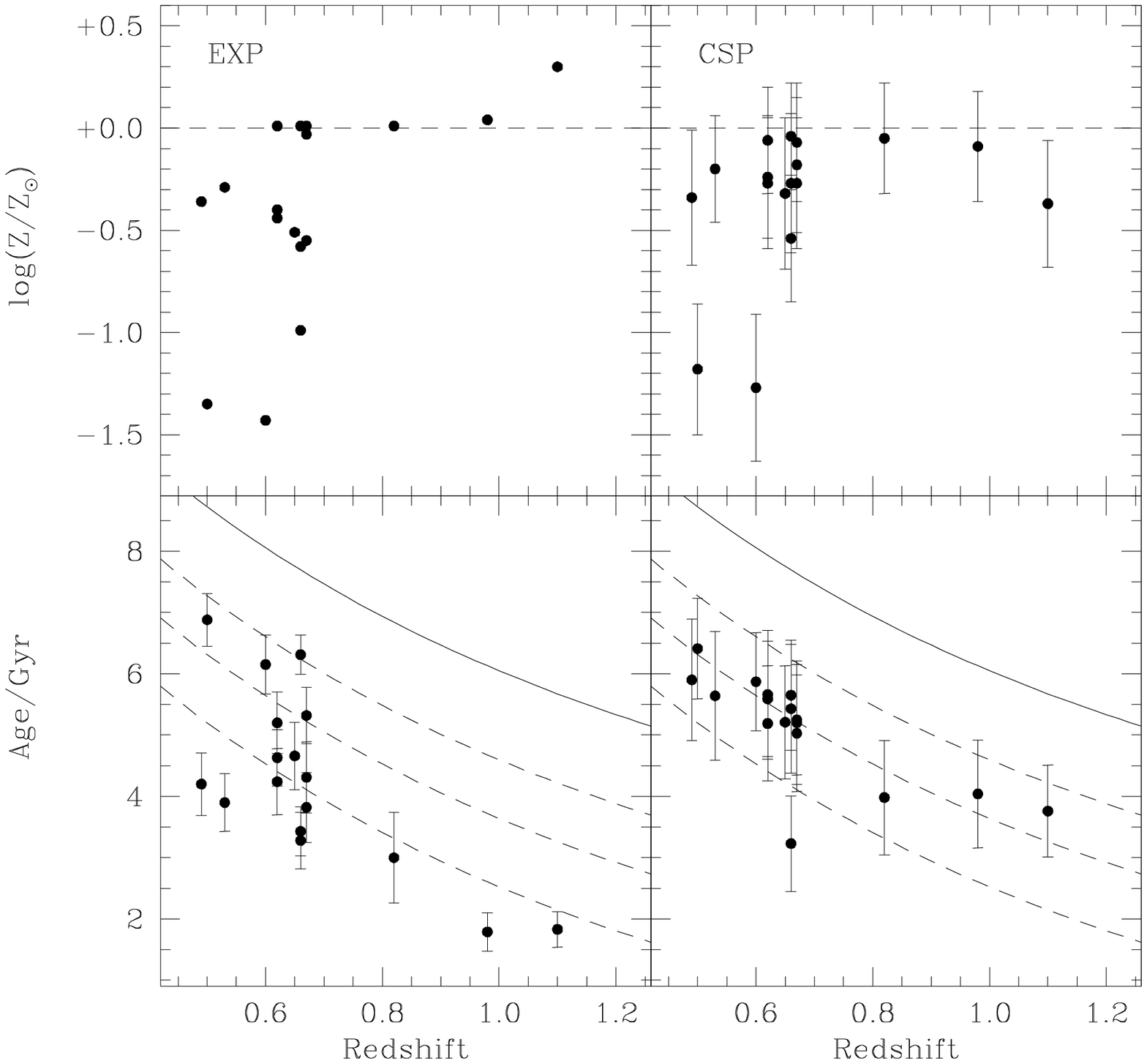}
\end{center}
FIG. 3.- Ages and metallicities corresponding to the best fit according to a simple
exponentially decaying model (EXP; left) or a consistent chemical enrichment code
(CSP; right). The dots are the average values of age or metallicity and the error bars
represent the RMS of the distribution. Notice the EXP models assume a fixed metallicity
for each star formation history. The solid lines in the bottom panel track the age of the
Universe at a given redshift for a concordance cosmology. The dashed lines -- from
top to bottom -- correspond to formation redshifts of $z_F=\{5,3,2\}$.
\label{fig:tZ}
\end{figure}

%\clearpage
\begin{figure}
\begin{center}
\includegraphics[width=3.8in]{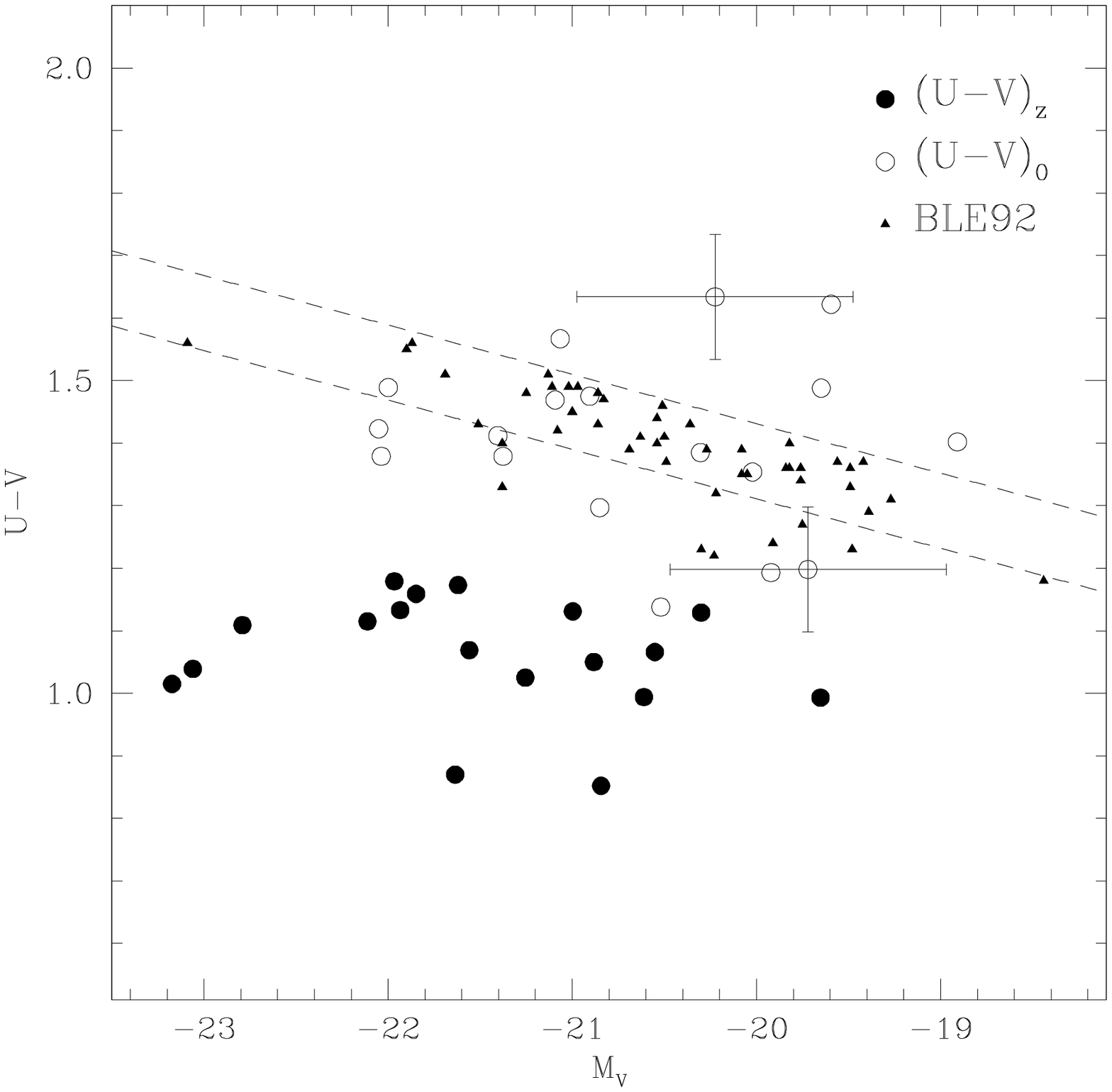}
\end{center}
FIG. 4.- Rest-frame $U-V$ vs $M_V$ color-magnitude relation. The filled dots correspond
to our sample after a K-correction to restframe $U$ and $V$ passbands. The hollow circles
are the $z=0$ color-magnitude relation of the same sample. In order to evolve the stellar
populations from their observed redshifts, we use the CSP model that gives the best fit.
Typical error bars (including the effect of the K-correction) are shown. The triangles
are Coma cluster galaxies from Bower et al. (1992) and the dashed lines delimit the
best fit and scatter of this local cluster.
\label{fig:CMR}
\end{figure}

\clearpage
\begin{figure}
\begin{center}
\includegraphics[width=3.8in]{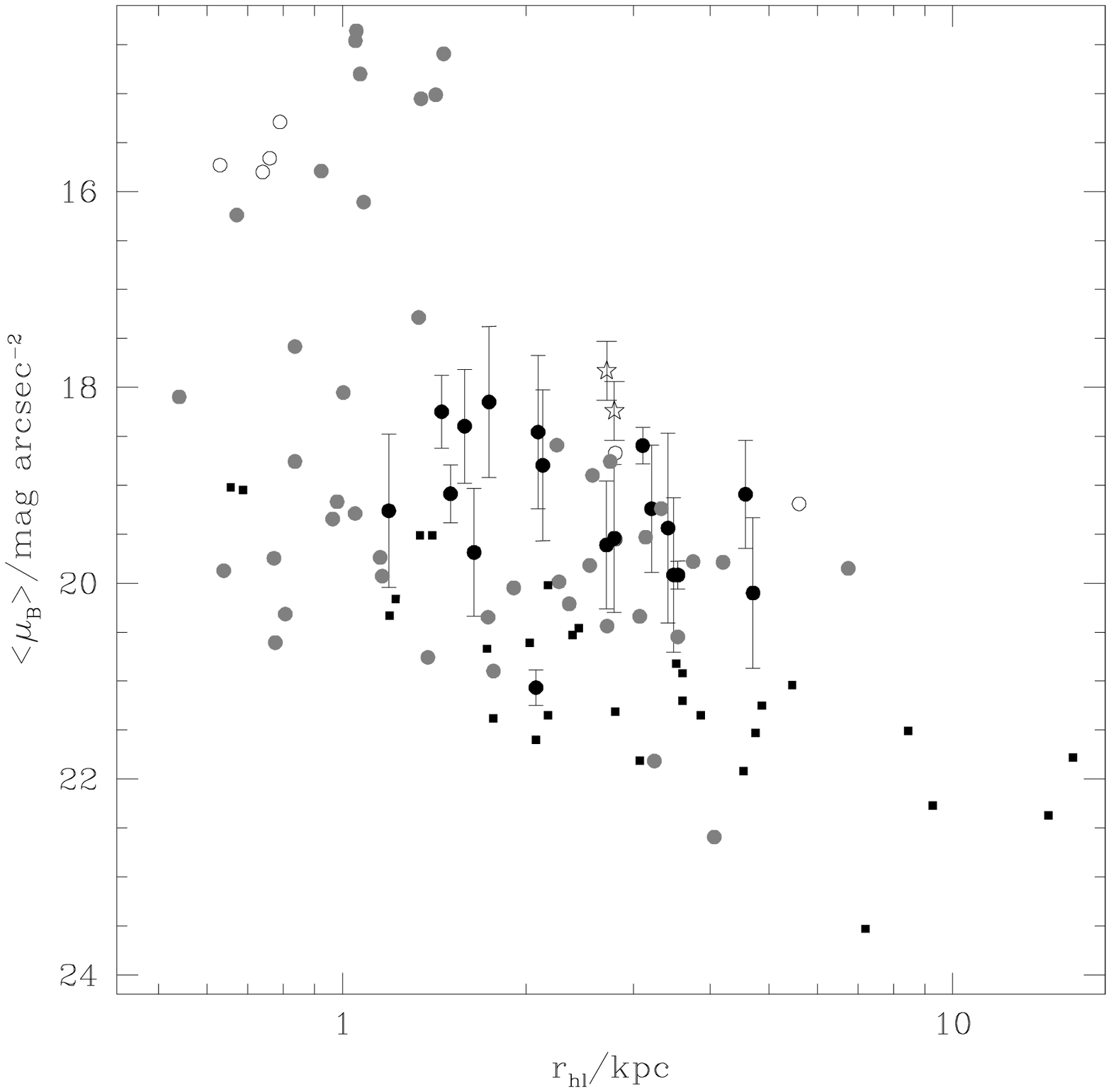}
\end{center}
FIG. 5.- $B$-band Kormendy relation of our sample (black dots) compared to Coma
cluster elliptical galaxies (squares; J\o rgensen et al. 1995) and 
early-type galaxies from the HDF (grey dots; Fasano et al. 1998).
The stars are the LBDS radio elliptical galaxies at $z\sim 1.5$
from Waddington et al. (2002), while the open dots represent the early-type
galaxies at $z \sim$ 2 studied by Daddi et al. (2005).
\label{fig:kor}
\end{figure}

%\clearpage
\begin{figure}
\begin{center}
\includegraphics[width=5.2in]{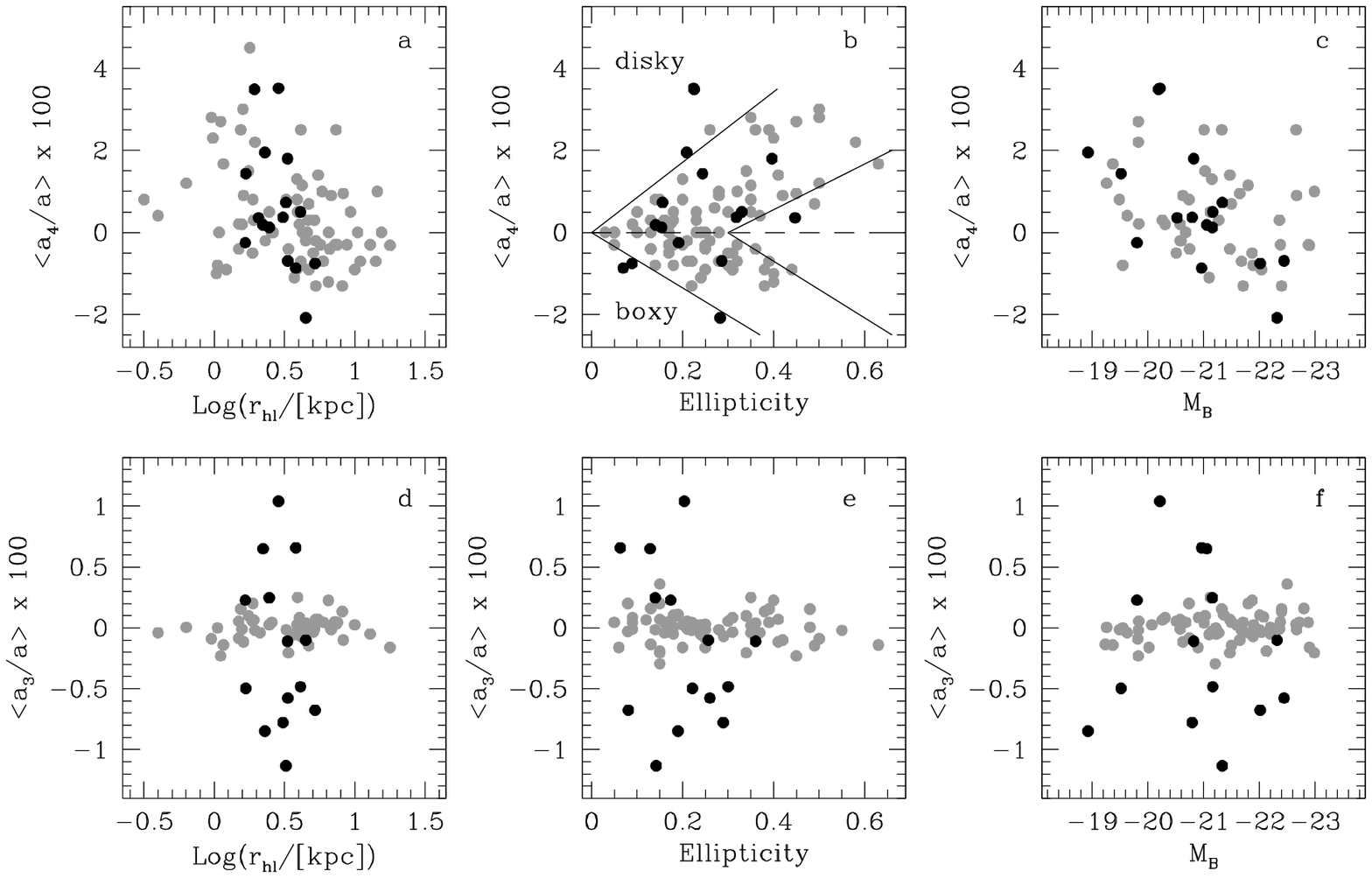}
\end{center}
FIG. 6.- The isophotal parameters of the {\it bona fide} early-type galaxies
(black, filled circles) and of the nearby ellipticals studied by Bender
et al. (1988, 1989, grey and filled circles). These parameters are also
plotted as a function of the galaxy half-light radius r$_{hl}$ and total absolute
magnitude in restframe B band M$_B$. The solid lines in panel b highlight the region
in ellipticity and {\it a$_4$/a} populated by nearby ellipticals. The ellipticity
and the {\it a$_4$/a} parameter of the {\it bona fide} early-type galaxies have
been corrected for the PSF, i.e. ellipticities have been increased by 10$\%$ and
{\it a$_4$/a} decreased by 0.15.
\label{fig:params}
\end{figure}

\clearpage
\begin{figure}
\begin{center}
\end{center}
\par\noindent
FIG. 7.- The (V-I) color maps of the {\it bona fide} early-type
galaxies. Available at http://www.stsci.edu/science/grapes/papers/pasquali05.
\end{figure}

\clearpage
\begin{figure}
\begin{center}
\end{center}
\par\noindent
FIG. 8.- As in Figure 7. Available at http://www.stsci.edu/science/grapes/papers/pasquali05.  
\label{fig:cmap2}
\end{figure}

\clearpage
\begin{figure}
\begin{center}
\end{center}
\par\noindent
FIG. 9.- {\it Top:} A close-up of the central region of J033244.09-274541.5
(left panel) and its extinction map (right panel) showing no presence of substructures.
{\it Middle:} A close-up of the central region of UDF2387  and the
corresponding extinction map, showing in white dust features in absorption.
{\it Bottom:} A close-up of the central region of UDF3677 and its extinction
map, with dust features in absorption. Close-ups are in units of counts/s
in the F775W band, extinction maps are in units of magnitude.
Available at http://www.stsci.edu/science/grapes/papers/pasquali05.
\label{fig:cmap3}
\end{figure}

\clearpage
\begin{figure}
\begin{center}
\end{center}
\par\noindent
FIG. 10.- As in Figure 9, now showing substructures in emission. UDF4527 and UDF6747
are characterized by disk-like features, while UDF153 and UDF5177 represent galaxies
with blue clumps and peculiar features, respectively.
Available at http://www.stsci.edu/science/grapes/papers/pasquali05.
\label{fig:cmap4}
\end{figure}

\clearpage
\begin{figure}
\begin{center}
\includegraphics[width=6.2in]{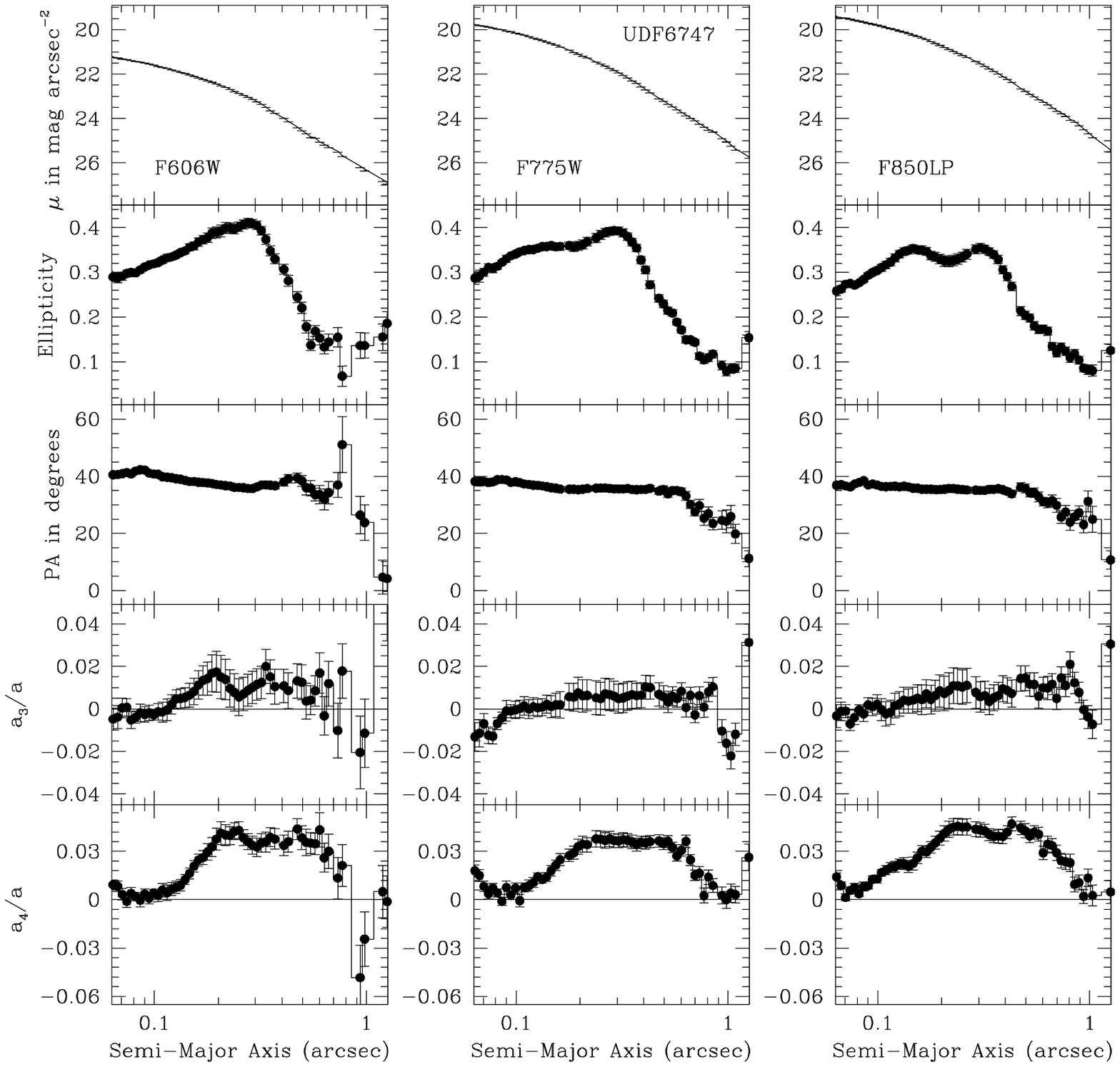}
\end{center}
\par\noindent
FIG. 11a.- The ELLIPSE parameters: surface brightness, ellipticity,
PA, {\it a$_3$/a}, {\it a$_4$/a} as a function of distance from the center of 
UDF6747 and derived in each of the F606W, F775W and F850LP filters.
This galaxy has a remarkable disk substructure (cf. Figure 10). Figures 11b - 11r
are available in the electronic edition of the Journal. The printed edition contains only
Figure 11a as an example.
\label{fig:ell1}
\end{figure}

\begin{figure}
\begin{center}
\includegraphics[width=4.2in]{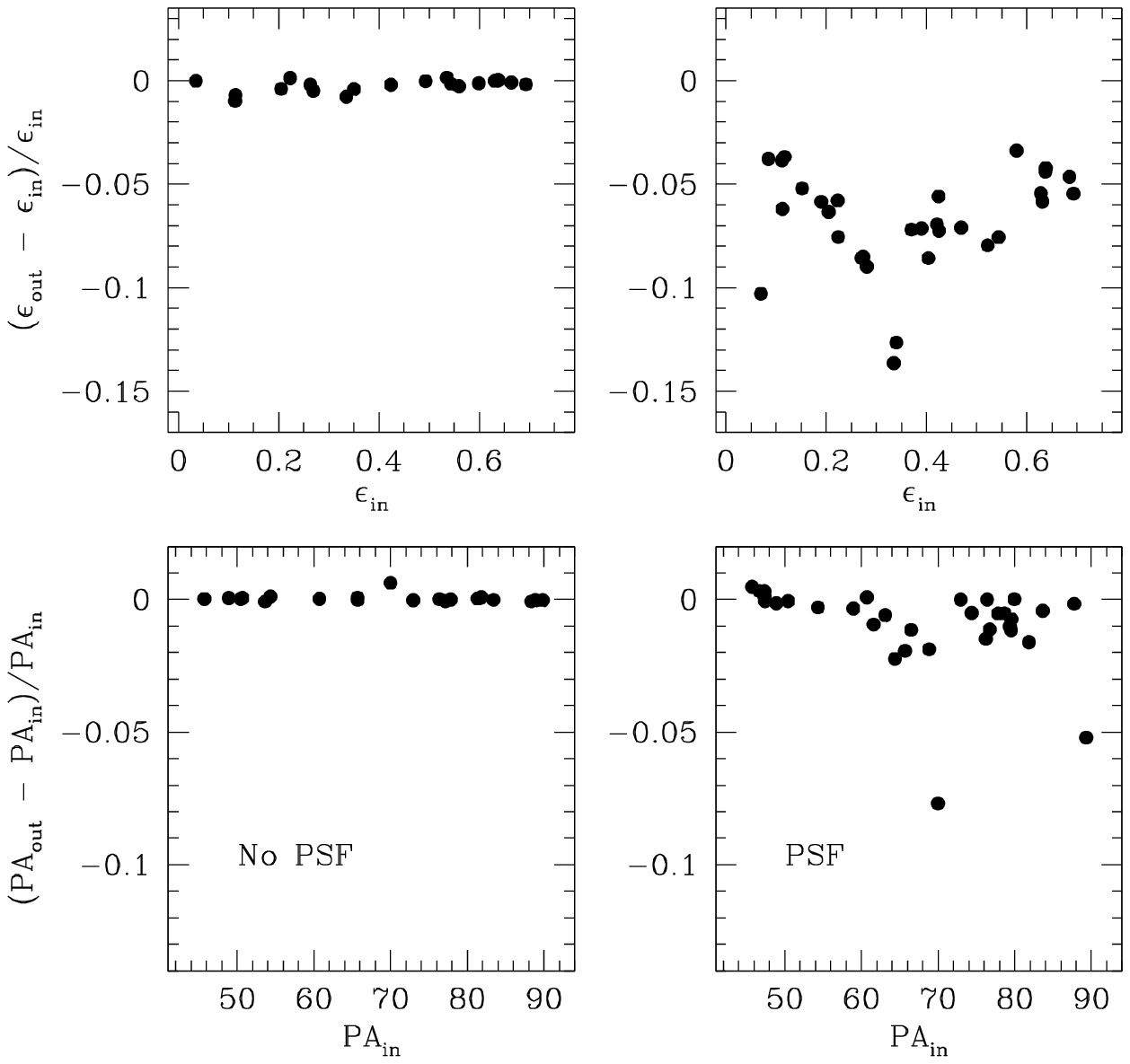}
\end{center}
\par\noindent
FIG. 12.- The comparison between the input and output ellipticities and PAs,
for the simulated galaxies with (right-hand side panels) and without 
(left-hand side panels) convolution with the ACS PSF in the F775W.
\label{fig:A1}
\end{figure}

\begin{figure}
\begin{center}
\includegraphics[width=4.2in]{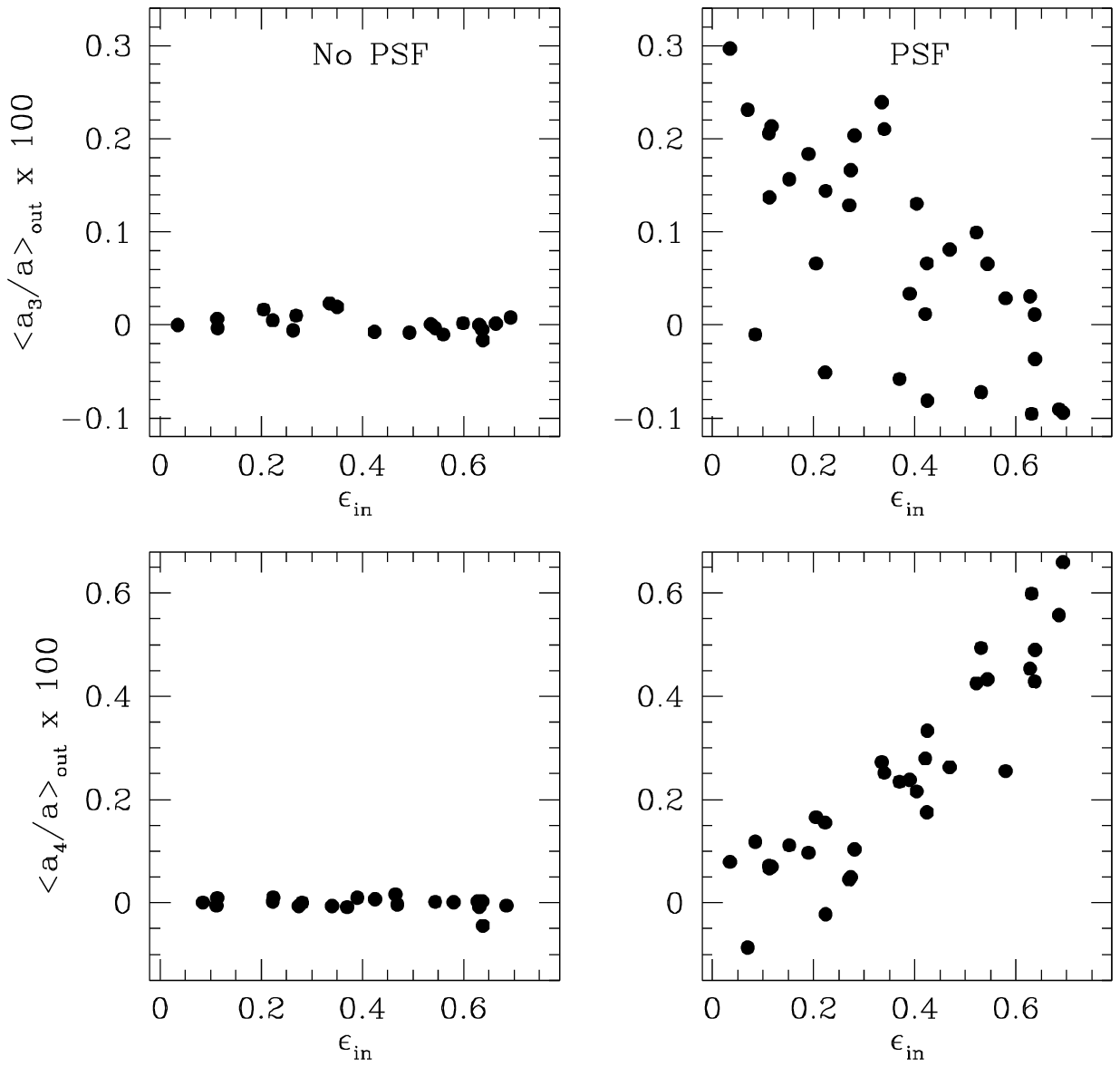}
\end{center}
\par\noindent
FIG. 13.- The comparison between the input and output {\it $\langle a_3/a\rangle$} and
{\it $\langle a_4/a\rangle$} for the simulated galaxies with (right-hand side panels)
and without (left-hand side panels) convolution with the ACS PSF in 
the F775W.
\label{fig:A2}
\end{figure}

\begin{thebibliography}{}
\bibitem[]{} Abraham, R.G., Ellis, R.S., Fabian, A.C. et al., 1999, MNRAS, 303, 641
\bibitem[~]{} Abraham, R.G., van den Bergh, S., Glazebrook, K. et al.,
1996, ApJS, 107, 1
%\bibitem[~]{} Adelberger, K.L., Steidel, C.C., Pettini, M., Shapley, A.E., Reddy, N.A.,
%Erb, D.K., 2005, ApJ, 619, 697
\bibitem[~]{} Alexander, D.M., Bauer, F.E., Brandt, W.N. et al., 2003, AJ, 126, 539
\bibitem[~]{} Barnes, J.E., 1992, ApJ, 393, 484
\bibitem[~]{} Beckwith, S.V.W. et al., 2005, in preparation
\bibitem[~]{} Bell, E.F., Wolf, C., Meisenheimer, K. et al., 2004, ApJ, 608, 752
\bibitem[~]{} Bender, R., D\"obereiner, S., M\"ollenhoff, C., 1988, A\&A Suppl.
Ser., 74, 385
\bibitem[~]{} Bender, R., Surma, P., D\"obereiner, S., M\"ollenhoff, C.,
Madejsky, R., 1989, A\&A, 217, 35
\bibitem[~]{} Bernardi, M., Sheth, R.K., James, A. et al., 2003, AJ, 125, 1882
\bibitem[~]{} Bershady, M.A., Jangren, J.A., Conselice, C.J., 2000, AJ, 119, 2645
\bibitem[~]{} Bertin, E. \& Arnouts, S., 1996, A\&AS, 117, 393
\bibitem[~]{} Bertoldi, F., Carilli, C.L., Menten, K.M. et al., 2000, A\&A, 360, 92
%\bibitem[~]{} Blaizot, J., Guiderdoni, B., Devriendt, J.E.G., Bouchet, F.R., Hatton, S.J.,
%Stoehr, F., 2004, MNRAS, 352, 571
\bibitem[~]{} Bolzonella, M., Miralles, J.-M., Pell\'o, R., 2000, A\&A, 363. 476
\bibitem[~]{} Bower, R.G., Lucey, J.R., Ellis, R.S., 1992, MNRAS, 254, 601
\bibitem[~]{} Brichmann, J., Ellis, R.S., 2000, ApJ, 536, L77
\bibitem[~]{} Bruzual, G. \& Charlot, S., 2003, MNRAS, 344, 1000
\bibitem[~]{} Caon, N., Capaccioli, M., D'Onofrio, M., 1993, MNRAS, 265, 1013
\bibitem[~]{} Cappellari, M., Copin, Y., 2003, MNRAS, 342, 345
\bibitem[~]{} Cimatti, A., Mignoli, M., Daddi, E. et al., 2002, A\&A, 392, 395 
\bibitem[~]{} Chapman, S.C., Blain, A.W., Ivison, R.J., Smail, I., 2003, Nature, 422, 695
\bibitem[~]{} Chapman, S.C., Windhorst, R.A., Odewahn, S., Yan, H., Conselice, C.J., 2004,
ApJ, 599, 92
\bibitem[~]{} Chen, H.-W., Marzke, R.O., McCarthy, P.J. et al., 2003, ApJ, 586, 745
\bibitem[~]{} Colley, W.N., Rhoads, J.E., Ostriker, J.P., Spergel, D.N., 1996, ApJL, 473, L63 
\bibitem[~]{} Conselice, C.J., 2003, ApJS, 147, 1
\bibitem[~]{} Conselice, C.J., Bershady, M.A., Dickinson, M., Papovich, C., 2003, AJ, 126, 1183
\bibitem[~]{} Conselice, C.J., Bershady, M.A., Jangren, J.A., 2000, ApJ, 529, 886
\bibitem[~]{} Conselice, C.J., Blackburne, J.A., Papovich, C., 2005, ApJ, 620, 564
\bibitem[~]{} Cross, N.J., Bouwens, R.J., Ben\'itez, N. et al., 2004, AJ, 128, 1990
\bibitem[~]{} Daddi, E., Renzini, A., Pirzkal, N. et al., 2005, ApJ, 626, 680 
%\bibitem{dav01} Davies, R.~L., Kuntschner, H., Emsellem, E. et al., 2001, ApJ, 548, L33
\bibitem[~]{} Dey, A., van Breugel, W.J.M., Vacca, W., Antonucci, R., 1997, ApJ, 490, 698
%\bibitem{sauron} de~Zeeuw, T., Bureau, M., Emsellem, E. et al., 2002, MNRAS, 329, 513
\bibitem[~]{} Di Tullio, G.A., 1978, A\&A, 62, L17
\bibitem[~]{} Di Tullio, G.A., 1979, A\&A Suppl. Ser., 37, 591
\bibitem[~]{} Dressler, A., 1980, 236, 351
\bibitem[~]{} Dressler, A., Oemler, Ag.Jr., Couch, W.J. et al., 1997, ApJ, 490, 577
\bibitem[~]{} Driver, S.P., Fernandez-Soto, A., Couch, W.J. et al., 1998, ApJ, 496, 93
\bibitem[~]{} Driver, S.P., Windhorst, R.A., Griffiths, R., 1995, ApJ, 453, 48
\bibitem[~]{} Ebneter, K., Balick, B., 1985, AJ, 90, 183
\bibitem[~]{} Eggen, O.J., Lynden-Bell, D., Sandage, A.R., 1962, ApJ, 136, 748
\bibitem[~]{} Fasano, G., Cristiani, S., Arnouts, S. 
  \& Filippi, M., 1998, AJ, 115, 1400
\bibitem[~]{} Fassnacht, C.D., Moustakas, L.A., Casertano, S. et al., 2004, ApJ, 600, L155
\bibitem[~]{} Ferreras, I., Lisker, T., Carollo, C.M., Lilly, S.J., Mobasher, B., 2005, astro-ph/0504127
\bibitem[~]{} Ferreras, I. \& Silk, J., 2000, MNRAS, 316, 786
\bibitem[~]{} Franx, M., 1988, MNRAS, 231, 285
\bibitem[~]{} Galletta, G., 1980, A\&A, 81, 179
\bibitem[~]{} Genzel, R., Baker, A.J., Tacconi, L.J. et al, 2003, ApJ, 584, 633
\bibitem[~]{} Giacconi, R., Zirm, A., Wang, J.X. et al., 2002, ApJS, 139, 369
\bibitem[~]{} Giavalisco, M., Ferguson, H.C., Koekemoer, A.M. et al., 2004, ApJ, 600, 93
\bibitem[~]{} Gilli, R., Cimatti, A., Daddi, E. et al., 2003, ApJ, 592, 721
\bibitem[~]{} Glazebrook, K., Ellis, R., Santiago, B., Griffiths, R., 1995, MNRAS, 275, L19
\bibitem[~]{} Goto, T., 2005, MNRAS, 357, 937 
\bibitem[~]{} Graham, A., Erwin, P., Caon, N., Trujillo, I., 2001, ApJ, 563, L11
\bibitem[~]{} Graham, A., Lauer, T.R., Colless, M., Postman, M., 1996, ApJ, 465, 534
\bibitem[~]{} Hamilton, D., 1985, ApJ, 297, 371
\bibitem[~]{} Hernquist, L., 1992, ApJ, 400, 460
\bibitem[~]{} Jedrzejewski, R.I., 1987, MNRAS, 226, 747
\bibitem[~]{} Jesseit, R., Naab, Y., Burkert, A., 2005, submitted to MNRAS, astro-ph/0501418
\bibitem[~]{} J\o rgensen, I., Franx, M. \& Kj\ae rgaard, P., 1995, MNRAS, 273 1097
\bibitem[~]{} Kauffmann, G. \& Charlot, S., 1998, MNRAS, 297, L23
\bibitem[~]{} Kauffmann, G., White, S.D.M., Guiderdoni, B., 1993, MNRAS, 264, 201
\bibitem[~]{} Kelson, D.D., Illingworth, G.D., van Dokkum, P.G., Franx, M., 2000, ApJ, 531, 184
\bibitem[~]{} Khochfar, S., Burkert, A., 2005, MNRAS, 359, 1379 
\bibitem[~]{} Kormendy, J., 1977, ApJ, 218, 333
\bibitem[~]{} Lauer, T.R., 1985, MNRAS, 216, 429
\bibitem[~]{} Lauer, T.R., Faber, S.M., Gebhardt, K. et al., 2005, AJ, 129, 2138
\bibitem[~]{} Le~F\`evre, O., Guzzo, L., Meneux, B. et al., 2004, A\&A, astro-ph/0403628
%\bibitem{lon89} Longo, G., Busarello, G., Capaccioli, M., Bender, R., 1989,
%A\&A, 225, 17
\bibitem[~]{} Malin, D.F., Carter, D., 1983, ApJ, 274, 534
\bibitem[~]{} Martel, A.R., Ford, H.C., Bradley, L.D. et al., 2004, AJ, 128, 2758 
\bibitem[~]{} Menanteau, F., Abraham, R.~G. \& Ellis, R.~S., 2001, MNRAS, 322, 1
\bibitem[~]{} Menanteau, F., Martel, A.R., Tozzi, P. et al., 2005, ApJ, 620, 697 
\bibitem[~]{} Michard, R., Poulain, P., 2000, A\&AS, 141, 1
\bibitem[~]{} Michard, R., Prugniel, P., 2004, A\&A, 423, 833
\bibitem[~]{} Mobasher, B., Idzi, R., Ben\'itez, N. et al., 2004, ApJ, 600, L167
\bibitem[~]{} Moustakas, L.A., Casertano, S., Conselice, C.J. et al., 2004, ApJ, 600, 131
\bibitem[~]{} Naab, T., Burkert, A., 2003, ApJ, 597, 893
\bibitem[~]{} Naab, T., Burkert, A. \& Hernquist, L., 1999, ApJ, 523, L133
\bibitem[~]{} Nelan, J.E., Smith, R.J., Hudson, M.J. et al., 2005, ApJ, accepted (astro-ph/0505301)
\bibitem[~]{} O'Dea, C.P., Koekemoer, A.M., Baum, S.A. et al., 2001, ApJ, 121, 1915
\bibitem[~]{} Odewahn, S.C., Burstein, D., Windhorst, R.A., 1997, AJ, 114, 2219
\bibitem[~]{} Papovich, C., Dickinson, M., Ferguson, H.C., 2001, ApJ, 559, 620
\bibitem[~]{} Papovich, C., Giavalisco, M., Dickinson, M. et al., 2003, ApJ,
598, 827
\bibitem[~]{} Peletier, R.F., Balcells, M., 1996, AJ, 111, 2238
\bibitem[~]{} Pettini, M., Shapley, A.E., Steidel, C.C. et al., 2001, ApJ, 554, 981
\bibitem[~]{} Pirzkal, N., Xu, C., Malhotra, S. et al., 2004, ApJS, 154, 501
\bibitem[~]{} Quinn, P.J., 1984, ApJ, 279, 596
\bibitem[~]{} Rest, A., van den Bosch, F.C., Jaffe, W. et al., 2001, AJ, 121, 2431
\bibitem[~]{} Sadler, E.M., Gerhard, O.E., 1985, MNRAS, 214, 177
\bibitem[~]{} S\`ersic, J.-L., 1968, Atlas de Galaxias Australes (Cordoba:
Obs. Astron.)
\bibitem[~]{} Schade, D., Lilly, S.J., Crampton, D. et al., 1999, ApJ, 525, 31
\bibitem[~]{} Shapley, A.E., Steidel, C.C., Adelberger, K.L., Dickinson, M., Giavalisco,
M., Pettini, M., 2001, ApJ, 562, 95
\bibitem[~]{} Smail, I., Ivison, R.J., Blain, A.W., Kneib, J.-P., 2002, MNRAS, 331, 495
\bibitem[~]{} Stanford, S.~A., Eisenhardt, P.~R.  \& Dickinson, M. 1998, ApJ, 492, 461
\bibitem[~]{} Steidel, C.C., Giavalisco, M., Pettini, M., Dickinson, M., Adelberger, K.L.,
1996, ApJ, 462, L17
\bibitem[~]{} Thielemann, F.-K., Nomoto, K. \& Hashimoto, M., 1996, ApJ, 460, 408
\bibitem[~]{} Thomas, D., 1999, MNRAS, 306, 655
\bibitem[~]{} Thomas, D., Greggio, L. \& Bender, R., 1999, MNRAS, 302, 537
\bibitem[~]{} Thomas, D., Maraston, C., Bender, R., de Oliveira Mendes, C., 2005, ApJ, 621, 673
\bibitem[~]{} Toomre, A., Toomre, J., 1972, ApJ, 178, 623
\bibitem[~]{} Trager, S.C., Faber, S.M., Worthey, G., Gonz\'alez, J.J., 2000, AJ, 119, 1645
\bibitem[~]{} Tran, H.D., Tsvetanov, Z., Ford, H.C. et al., 2001, AJ, 121, 2928
\bibitem[~]{} Treu, T., Ellis, R.S., Liao, T.X. et al., 2005, ApJ, submitted, astro-ph/0503164
\bibitem[~]{} van Breugel, W.J.M., Stanford, S.A., Spinrad, H., Stern, D., Graham, J.R., 1998,
ApJ, 502, 614
\bibitem[~]{} van den Bosch, F.C., Ferrarese, L., Jaffe, W., Ford, H.C., O'Connell,
R.W., 1994, AJ, 108, 1579
\bibitem[~]{} van~den~Hoek, L.~B. \& Groenewegen, M.~A.~T., 1997, A\&AS, 123, 305
\bibitem[~]{} van Dokkum, P.G., Franx, M., 1995, AJ, 110, 2027
\bibitem[~]{} van Dokkum, P.G., Franx, M., 2001, ApJ, 553, 90
\bibitem[~]{} van Dokkum, P.G., Franx, M., Kelson, D.D., Illingworth, G.D., 1998, ApJ, 504, L17
\bibitem[~]{} Vanzella, E., Cristiani, S., Dickinson, M. et al., 2005, A\&A, 434, 53 
\bibitem[~]{} Yang, Y., Zabludoff, A.I., Zaritsky, D., Lauer, T.R., Mihos, J.C.,
2004, ApJ, 607, 258 
\bibitem[~]{} Waddington, I., Windhorst, R.A., Cohen, S.H. et al., 2002, MNRAS, 336, 1342
\bibitem[~]{} Webb, T.M.A., Eales, S.A., Lilly, S.J. et al., 2003, ApJ, 587, 41
\bibitem[~]{} Williams, R.E., Blacker, B., Dickinson, M. et al., 1996, AJ, 112, 1335
\bibitem[~]{} Windhorst, R.A., Taylor, V.A., Jansen, R.A. et al., 2002, ApJS, 143, 113
\bibitem[~]{} Worthey, G., 1994, ApJS, 95, 107
\end{thebibliography}
\end{document}